\documentclass[twocolumn,prd,preprintnumbers,amsmath,amssymb,nofootinbib,superscriptaddress]{revtex4-2}
\usepackage[utf8]{inputenc}
\usepackage[dvipsnames]{xcolor}
\usepackage{amsmath}
\usepackage{makeidx} 
\usepackage{amssymb}
\usepackage{graphicx}
\newcommand{\ph}{\phantom}
\newcommand{\nn}{\nonumber}

\newcommand{\eps}{\epsilon}

\newcommand{\vareps}{\varepsilon}

\newcommand{\dpo}{D^{({\cal P})}}
\newcommand{\dlo}{D^{({\cal A})}}
\newcommand{\dlosd}{D^{({\cal A}^{+})}}

\newcommand{\calx}{{\cal X}}
\newcommand{\idxI}{\scriptscriptstyle{I}}
\newcommand{\idxJ}{\scriptscriptstyle{J}}
\newcommand{\idxK}{\scriptscriptstyle{K}}
\newcommand{\idxL}{\scriptscriptstyle{L}}

\usepackage{tipa}

\begin{document}


\title{Paths to gravitation via the gauging of parameterized field theories}

\author{Tomi Koivisto}
\email{tomi.koivisto@ut.ee}
\affiliation{Laboratory of Theoretical Physics, Institute of Physics,
	University of Tartu, W. Ostwaldi 1, 50411 Tartu, Estonia and
	National Institute of Chemical Physics and Biophysics, R{\"a}vala pst. 10, 10143 Tallinn, Estonia}                           
\author{Tom Zlosnik}
\email{thomas.zlosnik@ug.edu.pl}
\affiliation{University Of Gda\'{n}sk, Jana Ba{\.z}y\'{n}skiego 8, 80-309 Gda\'{n}sk, Poland}             
                                                                  
\date{\today}
	
\begin{abstract}
	In special-relativistic physics, spacetime is imbued with a fixed, non-dynamical metric tensor. A path to gravitational theory is to promote this tensor to a genuine dynamical field. An alternative description of special-relativistic physics involves no fixed spacetime geometry but instead the inclusion of scalar fields $X^{I}(x^{\mu})$ which dynamically may take the form of inertial coordinates in spacetime. This suggests an alternative approach to gravity where the invariance of actions under global Poincar\'{e} transformations of $X^{I}$ is promoted to either a local Poincar\'{e}, local translational, or local Lorentz symmetry via the introduction of gauge fields. Points of commonality and departure of the resulting gravitational theories as compared to General Relativity are discussed. It is shown that the model based on local Lorentz symmetry is an extension of General Relativity that can introduce a standard of time into the dynamics of the gravitational field and allows for spacetimes described by a Minkowski metric or flat Euclidean signature metric despite the gravitational gauge field possessing non-zero curvature.
\end{abstract}

\maketitle

\section{Introduction}
The notion of gauge symmetry is a crucial part of the mathematical structure of the standard model of particle physics. Consider an action $S_{\chi}[\chi]$ describing the dynamics of a matter field $\chi$ that is invariant under a global (i.e. independent of location in spacetime) continuous symmetry represented by the transformation $\chi \rightarrow U\chi$ (where indices are suppressed for notational compactness). Typically the action will not be invariant under local symmetry transformations ($U=U(x^{\mu})$) as derivative terms $\partial_{\mu}\chi$ present in the Lagrangian then do not transform homogeneously under this transformation. However, the global symmetry can generally be promoted to a local one by the introduction of an additional field $A_{\mu}$ - called a gauge field or connection - which allows for the creation of a \emph{covariant derivative} which - if the transformation $U$ can be represented as a matrix and $\chi$ belongs to the fundamental representation of the symmetry group - takes the form:

\begin{align}
\partial_{\mu}\chi \rightarrow D^{(A)}_{\mu}\chi \equiv \partial_{\mu}\chi + A_{\mu}\chi
\end{align}
If, under the $U$ transformation, $A_{\mu} \rightarrow UA_{\mu}U^{-1} - \partial_{\mu}U U^{-1}$ then $D^{(A)}_{\mu}\chi \rightarrow UD^{(A)}_{\mu}\chi$. The extension of the definition of $D^{(A)}_{\mu}$ to matter fields in other representations of the symmetry group is straightforward. Alongside the modification $S_{\chi}[\chi]\rightarrow S_{\chi}[\chi,A_{\mu}]$, the process is then completed by the introduction of an action $S_{A}[A_{\mu},\chi]$ which allows for the dynamics of $A_{\mu}$ to be well defined. An example of this process would be that of a complex scalar field theory, where the Lagrangian density in inertial coordinates in Minkowski spacetime is ${\cal L}_{\phi} =-\frac{1}{2}\eta^{\mu\nu}\partial_{\mu}\phi^{*}\partial_{\nu}\phi - V(\phi^{*}\phi)$ is invariant under global $U(1)$ transformations $\phi \rightarrow e^{i\alpha}\phi$. The $U(1)$ invariance can be made local by introducing a field $A_{\mu}$ to construct the covariant derivative $D_{\mu}^{(A)}\phi =\partial_{\mu}\phi + A_{\mu}\phi$; the resultant locally $U(1)$ invariant action for $\phi$ is then supplemented by the Lagrangian density ${\cal L}_{A} = -\frac{1}{4}F^{\mu\nu}F_{\mu\nu}$, where $F_{\mu\nu} = 2\partial_{[\mu}A_{\nu]}$, which provides dynamics for the field $A_{\mu}$.

In this paper we will consider a similar gauging process in the context of gravitation, first reviewing existing results that show how it can be used to recover General Relativity then exploring a new variant that yields novel gravitational dynamics. 

Special relativistic theories are commonly formulated in terms of matter fields existing in a space with fixed-geometrical structure i.e. Minkowski space and its accompanying metric tensor $\eta_{\mu\nu}$. An alternative approach is to not assume the presence of $\eta_{\mu\nu}$ but rather introduce a set of four scalar fields $X^{I}(x^{\mu})$ which are dynamical in the sense the action is stationary with respect to small variations of these fields and this results in them having own equations of motion. Actions can be constructed so that dynamically the fields end up configured to play the role of inertial coordinate fields in spacetime with an effective metric emerging via the combination $
\tilde{\eta}_{\mu\nu} =\eta_{IJ}\partial_{\mu}X^{I}\partial_{\nu}X^{J}$, where $\eta_{IJ}=\mathrm{diag}(-1,1,1,1)$. Such theories are referred to as `parameterized field theories'. An interesting property of these theories is that the absence of fixed geometrical structure (such as $\eta_{\mu\nu}$) means that such actions possess a symmetry with respect to spacetime diffeomorphisms in the manner familiar from gravitational theory. In addition, the actions for parameterized field theories possess a global symmetry:

\begin{align}
X^{I} \rightarrow \Lambda^{I}_{\ph{I}J}X^{J} + P^{I} \label{globpoin}
\end{align}
The combined effect of the orthogonal matrix $\Lambda^{I}_{\ph{I}J}$ (representing a Lorentz transformation of $X^{I}$) with $P^{I}$ is that of a global Poincar\'{e} transformation of $X^{I}$, analogous to the global coordinate transformations that preserve $\eta_{\mu\nu} = (-1,1,1,1)$ i.e. that preserve the form of the Minkowski metric in inertial coordinates.

It will be shown that the promotion of the global symmetry (\ref{globpoin}) to a local one via the introduction of gauge fields leads to gravitational theory. Specifically, if the entire transformation (\ref{globpoin}) is promoted to a local one then the Einstein-Cartan theory of gravity is recovered (see, for example \cite{Grignani:1991nj}). We then discuss the result that instead promoting \emph{only} the symmetry under global translations of $X^{I}$ to a local one results in the teleparallel formulation of General Relativity. 
These first two examples are known in the literature. 
We will then show that a novel extension to General Relativity can be recovered if, instead, only the global Lorentz symmetry is is promoted to a local symmetry, alongside the removal of the global invariance under $X^{I} \rightarrow X^{I} + P^{I}$. 

The structure of the paper is as follows. In Section \ref{section_parameterization} we introduce the notion of parameterization, beginning with parameterized particle mechanics where models of mechanics are formulated in a way such that Newtonian time is promoted to an independent, dynamical field and then proceeding to parameterized field theory in four dimensions with the introduction of four dynamical `coordinate' fields $X^{I}$. Section  \ref{section_gaugingPoincare} contains an overview of existing results in the literature relating gravity to notions of the gauging of global Poincar\'{e} symmetry: specifically, Kibble's gauging of the global Poincar\'{e} symmetry present in non-parameterized field theory and of how the gauging of the global Poincar\'{e} symmetry that the $X^{I}$ fields possess in parameterized field theories leads to the first-order Einstein-Cartan theory of gravity. 
Section \ref{section_gaugingTranslations} continues the survey of  existing results with the demonstration of how gauging \emph{only} the global translations of $X^{I}$ can lead to the teleparallel formulation of General Relativity. 
In Section \ref{section_gaugingLorentz} we present a new possibility: that, instead, gauging only the global Lorentz symmetry that the $X^{I}$ fields possess leads to an extension to General Relativity with novel phenomenology. In Section \ref{frw} we obtain solutions to this model in Friedmann-Robertson-Walker (FRW) symmetry, demonstrating the appearance of an additional `dark' matter density in the cosmological equations of motion. In Section \ref{minkowski} we obtain solutions to the model which correspond to Minkowski space; interestingly, there are two structurally distinct possibilities: one where the curvature of the gravitational gauge field is zero, one where it is \emph{non-zero}. The behaviour of small perturbations around these backgrounds is discussed. In Section \ref{other} we briefly discuss consequences of the complex-valuedness of gravitational fields in the model, such as the recovery of flat, four dimensional Euclidean space as another solution to the model's field equations. In Section \ref{section_phenomenology}
we discuss the phenomenology of the model and future steps that to enable comparison between theory and observation.
In \ref{discussandconclude} we discuss the collected results in the paper and present our conclusions.

\section{Parameterization}
\label{section_parameterization}

Newton's laws of motion describing a particle with position $q^{i}$ ($i=1,2,3$) follow from the stationarity of the following action under small variations of $q^{i}$:

\begin{align}
	S[q^{i}] = \int dT \bigg(\frac{m}{2}\delta_{ij}\frac{d q^{i}}{dT}\frac{d q^{j}}{dT}-V(q^{i})\bigg) \label{newtact}
\end{align}
The functional derivative $\delta S/\delta q^{i}=0$ is equivalent to:

\begin{align}
	m \frac{d}{dT}\bigg(\delta_{ij}\frac{dq^{j}}{dT}\bigg) = -\frac{\partial V}{\partial q^{i}}
\end{align}
Alternatively, one can look to promote the Newtonian time $T$ to an independent degree of freedom $T(\lambda)$ (alongside  $q^{i}=q^{i}(\lambda)$) where $\lambda$ parameterizes trajectories. We can consider the following action:

\begin{align}
	S'[q^{i},T] = \int d\lambda \bigg(\frac{dT}{d\lambda}\bigg) \bigg(\bigg(\frac{dT}{d\lambda}\bigg)^{-2}\frac{m}{2}\delta_{ij}\frac{d q^{i}}{d\lambda}\frac{d q^{j}}{d\lambda}-V(q^{i})\bigg) \label{ppm}
\end{align}
This is the action for parameterized particle mechanics and is invariant under reparameterizations $\lambda \rightarrow f(\lambda)$ that reduce to the identity ($\lambda \rightarrow \lambda$) at the end points of integration of the action. The first equation of motion $\delta S/\delta q^{i}=0$ is:

\begin{align}
	m\frac{d}{d\lambda}\bigg(\bigg(\frac{dT}{d\lambda}\bigg)^{-2}\delta_{ij}\frac{dq^{j}}{d\lambda}\bigg) &=  -\frac{\partial V}{\partial q^{i}}
\end{align}
whereas there is now a new equation of motion following from $\delta S/\delta T=0$:

\begin{align}
	\frac{d}{d\lambda}\bigg(\bigg(\frac{dT}{d\lambda}\bigg)^{-2}\frac{m}{2}\delta_{ij}\frac{d q^{i}}{d\lambda}\frac{d q^{j}}{d\lambda}+V(q^{i})\bigg) &=0
\end{align}
This equation can be integrated to yield:

\begin{align}
\bigg(\frac{dT}{d\lambda}\bigg)^{-2}\frac{m}{2}\delta_{ij}\frac{d q^{i}}{d\lambda}\frac{d q^{j}}{d\lambda}+V(q^{i}) &=E
\end{align}
where $E$ is a constant. There exist solutions where $T$ varies monotonically with $\lambda$ in which case a gauge/parameterization $\lambda \overset{*}{=} T$ can be found and the collective equations of motion are:

\begin{align}
	m\frac{d}{dT}\bigg(\delta_{ij}\frac{dq^{j}}{dT}\bigg) &\overset{*}{=}  -\frac{\partial V}{\partial q^{i}} \\
\frac{m}{2}\delta_{ij}\frac{d q^{i}}{dT}\frac{d q^{j}}{dT}+V(q^{i}) & \overset{*}{=} E
\end{align}
Therefore, Newton's equations of motion with solutions corresponding to a single value of energy $E$ are recovered. The theory also permits solutions where $T$ does not vary monotonically with $\lambda$, for example admitting solutions such as $\frac{dT}{d\lambda}=0$ where time seems not to flow or solutions where the sign of $\frac{dT}{d\lambda}$ varies. For such situations the gauge $T = \lambda$ is not globally accessible. Nonetheless, the quantum mechanical propagator for this theory can be constructed in terms of gauge/parameterization independent observables and reproduce the results of standard quantum mechanics based on the action (\ref{newtact}) with $T$ playing the role of time \cite{Henneaux:1992ig,Gryb:2008rz}.

Now we consider the extension of these ideas to field theory. For concreteness we consider the case of the electromagnetic field. The following action yields Maxwell's equations upon small variations of the field $A_{\mu}$:

\begin{align}
S'[A_{\mu}] &=  -\frac{1}{4} \int d^{4}x \sqrt{-\mathrm{det}[\eta]} \eta^{\mu\alpha}\eta^{\nu\beta}F_{\mu\nu}F_{\alpha\beta} \label{sprimeEM}
\end{align}
where $\eta_{\mu\nu}$ is the metric tensor of Minkowski spacetime,  $F_{\mu\nu} \equiv 2\partial_{[\mu} A_{\nu]}$, and $\{x^{\mu}\}$ are some set of coordinates describing points in spacetime (not necessarily Minkowski coordinates).
Due to the fixed, flat geometry of spacetime there exist `inertial' coordinate systems coordinatized by $\{X^{I}\}$, for which in a general coordinate system $\{x^{\mu}\}$

\begin{align}
\eta_{\mu\nu} &= \eta_{IJ}\frac{\partial X^{I}}{\partial x^{\mu}}\frac{\partial X^{J}}{\partial x^{\nu}}
\end{align}
where $\eta_{IJ} = \mathrm{diag}(-1,1,1,1)$. In the case of mechanics when described by the action (\ref{newtact}), Newtonian time $T$ appears as a non-dynamical, monotonically increasing parameter. From a modern perspective $T$ is, rather, reflective of the spacetime structure provided by the metric tensor $g_{\mu\nu}$, itself a dynamical field. One can imagine the motivation for promoting $T$ to a field to be independently varied and similarly, one might imagine looking to recover metric structure from  $\eta_{IJ}\frac{\partial X^{I}}{\partial x^{\mu}}\frac{\partial X^{J}}{\partial x^{\nu}}$ instead of the fixed background metric $\eta_{\mu\nu}$ and promote the fields $X^{I}$ to being dynamical. Then, consider the following action:

\begin{align}
	S'[A_{\mu},X^{I}] &=  -\frac{1}{4} \int d^{4}x \sqrt{-\mathrm{det}[\tilde{\eta}]} \tilde{\eta}^{\mu\alpha}\tilde{\eta}^{\nu\beta}F_{\mu\nu}F_{\alpha\beta} \label{sprimeEM}
\end{align}
where

\begin{align}
	\tilde{\eta}_{\mu\nu} &\equiv \eta_{IJ}\frac{\partial X^{I}}{\partial x^{\mu}}\frac{\partial X^{J}}{\partial x^{\nu}}
\end{align}
and where $\eta_{IJ} = \mathrm{diag}(-1,1,1,1)$ and $\tilde{\eta}^{\mu\nu}$ is the matrix inverse of $\tilde{\eta}_{\mu\nu}$, which is assumed to exist. The action (\ref{sprimeEM}) is manifestly invariant under the global transformation (\ref{globpoin}) and the equation of motion for $X^{I}$ can be shown to be

\begin{align}
0 &=  \partial_{\mu}\bigg(\sqrt{-\mathrm{det}[\tilde{\eta}]}\frac{\partial X_{I}}{\partial x^{\nu}}T^{\mu\nu}\bigg)
\end{align}
where $T_{\mu\nu}$ is the stress energy tensor of the electromagnetic field; therefore the equation of motion for $X^{I}$ expresses conservation of the stress energy tensor, analogously to the equation of motion for Newtonian time $T$ recovered the conservation of energy. There exist solutions $X^{I}(x^{\mu})$ for which there exist coordinates such that $\partial X^{I}/\partial x^{\mu} \overset{*}{=} \delta^{I}_{\mu}$. This is a generalization of the $\lambda \overset{*}{=}T$ gauge in parameterized particle mechanics. In these coordinates, $\tilde{\eta}_{\mu\nu} = \mathrm{diag}(-1,1,1,1)$ and so we see that $X^{I}$ here play the role of inertial coordinates in Minkowski spacetime. As expected, for these solutions, this form of $\tilde{\eta}_{\mu\nu}$ is preserved by the transformation (\ref{globpoin}). Generally, the recovery of familiar classical field theory in Minkowski space is possible in the parameterized approach, though interestingly the recovery of standard results in quantum field theory from the parameterized approach in four dimensions encounters a number of technical challenges \cite{Torre:1997zs}.

\section{Gravity via gauging of global Poincar\'{e} invariance}
\label{section_gaugingPoincare}
The standard route to gravitation has been via Einstein's General Relativity where $\eta_{\mu\nu}$ is promoted to a dynamical field (denoted $g_{\mu\nu}$) with its own action which is given - up to the necessary Gibbons–Hawking–York boundary term - by the Einstein-Hilbert action:

\begin{align} \label{einstein-hilbert}
	S_{g}[g_{\mu\nu}] &=  \frac{1}{16\pi G} \int d^{4}x \sqrt{-g} R(g)
\end{align}
This approach must be modified somewhat when fermionic fields are present. The actions of the standard model of particle physics consists of the following dynamical fields: gauge fields $A_{\mu}$ (spacetime one-forms), the electroweak Higgs field $\phi$ (a spacetime scalar), and fermionic fields $\Psi_{A}$ and $\chi^{A'}$ (Weyl spinors i.e. spacetime scalars in the fundamental representations of $SL(2,C)$) alongside the non-dynamical object $\bar{e}^{I}_{\mu}$ (a spacetime one form in the fundamental representation of $SO(1,3)$ such that $\eta_{IJ}\bar{e}^{I}_{\mu}\bar{e}^{J}_{\nu}\equiv \eta_{\mu\nu}$). These actions are invariant under global $SL(2,C)$ transformations which act only on the Weyl spinors and on $\bar{e}^{I}_{\mu}$ via the group homomorphism between $SL(2,C)$ and $SO(1,3)$. The Lagrangian four forms ${\cal L}$ that are integrated to produce the actions of the standard model transform as differential forms under diffeomorphisms that act on \emph{both} dynamical and non-dynamical fields but also transform as forms under diffeomorphisms that act \emph{only} on the dynamical fields for diffeomorphisms generated by vector fields $\xi^{\mu}$ that satisfy $\pounds_{\xi}\eta_{\mu\nu}=0$ (where $\pounds_{\xi}$ denotes the Lie derivative) i.e. $\xi^{\mu}$ that satisfy this equation are the Killing vectors of Minkowski space. There are ten independent $\xi^{(i)\mu}$ and their commutator $[\xi^{(i)},\xi^{(j)}]$ satisfies the Lie algebra of the Poincar\'{e} group $ISO(1,3)$. In this sense the actions of the standard model possess a global $SL(2,C)$ symmetry and the Lagrangian forms exhibit a global $ISO(1,3)$ covariance. It was shown by Kibble \cite{Kibble:1961ba} (building on earlier work by Utiyama \cite{Utiyama:1956sy}) that the global $SL(2,C)$ symmetry could be promoted to a local one by the introduction of a gauge field $\omega$ valued in the Lie algebra of $SL(2,C)$ - such that the covariant derivative $D_{\mu}^{(\omega)}\chi^{A'} \equiv \partial_{\mu}\chi^{A'} +\omega^{A'}_{\ph{A'}B'\mu} \chi^{B'}$, where  $\omega^{A'}_{\ph{A'}B'\mu} = \frac{1}{8}\omega_{IJ\mu}(\bar{\sigma}^{I}\sigma^{J}-\bar{\sigma}^{J}\sigma^{I})^{A'}_{\ph{A'}B'}$,  transforms homogeneously under this transformation. Additionally, the remaining presence of non-dynamical, prior geometry was removed by the introduction of a dynamical field $e^{I}_{\mu}$ to appear in place of $\bar{e}^{I}_{\mu}$ - such that $ \eta_{IJ}e^{I}_{\mu}e^{J}_{\nu}\equiv g_{\mu\nu}$.  The introduction of the set of dynamical fields $\{e^{I}_{\mu},\omega^{A'}_{\ph{A}B'\mu}\}$ into the matter actions suggests that the gauging process should be completed by providing action allowing for a consistent dynamics of these degrees of freedom i.e. the introduction of gravitation as a dynamical interaction. A simple possibility is the following action:

\begin{align}
	S_{g}[\omega,e] 	&= \frac{1}{32\pi G} \int \eps_{IJKL}e^{I}\wedge e^{J} \wedge R^{KL}(\omega) \nn\\
&= \frac{1}{64\pi G}\int d^{4}x \vareps^{\mu\nu\alpha\beta} \eps_{IJKL}e^{I}_{\mu} e^{J}_{\nu}  R^{KL}_{\ph{KL}\alpha\beta}(\omega) \label{eincartan}
\end{align}
where $\vareps^{\mu\nu\alpha\beta}$ is the Levi-Civita 
\emph{density} and $R^{IJ}_{\ph{IJ}\mu\nu} \equiv 2\partial_{[\alpha}\omega^{IJ}_{\ph{IJ}\beta]} +2 \omega^{I}_{\ph{I}K[\alpha}\omega^{KJ}_{\ph{KJ}\beta]}$. The action  (\ref{eincartan}) is the Palatini action in the Einstein-Cartan formulation of gravity and its equations of motion are classically equivalent to General Relativity with an additional matter term quadratic in fermionic currents. 

This procedure can be interpreted as a combined gauging of  internal symmetries and the limited spacetime covariances of the original non-gravitational actions which leads to a theory of matter and gravity that possess a local internal $SL(2,C)$ symmetry and is generally covariant in the sense that the action is invariant under infinitesimal diffeomorphisms generated by vector fields $\zeta^{\mu}$ that vanish at the boundary of the action's integration that act on the dynamical fields $\chi$ as $\chi \rightarrow \chi + \pounds_{\zeta}\chi$ \cite{Harlow:2019yfa}. 

We note that the gauging procedure does not uniquely fix the gravitational action but rather suggests a family of potential actions, each of which must possess a symmetry under both local $SL(2,C)$ transformations and spacetime diffeomorphisms. Indeed, allowing the gravitational action to consist of terms up to quadratic order in $R^{KL}(\omega)$ and $T^{I}\equiv de^{I}+\omega^{IJ}\wedge e_{J}$ is the approach of Poincar\'{e} gauge theory which permits a wealth of interesting phenomenology \cite{Hehl:2013qga,Ho:2015ulu,Obukhov:2017pxa,Blagojevic:2018dpz,BeltranJimenez:2019hrm,Barker:2020elg,Barker:2020gcp}. 

Now we return to the parameterized approach, now as applied to the actions of the standard model. In the place of the non-dynamical field $\bar{e}^{I}_{\mu}$ we instead have $\partial_{\mu}X^{I}$. Due to the fields $X^{I}$ now being dynamical, the actions are generally covariant as well as possessing an additional invariance under the following global transformations:

\begin{align}
X^{I} \rightarrow \Lambda^{I}_{\ph{I}J}X^{J} + P^{I} \label{xiglob}
\end{align}
where $\Lambda_{JI} = \Lambda^{-1}_{IJ}$ (where indices have been lowered with $\eta_{IJ}$). We may consider what happens if some or all of the global invariance under the transformation (\ref{xiglob}) is promoted to a \emph{local} invariance \cite{Grignani:1991nj}.
Consider the possibility that $\{\Lambda^{I}_{\ph{I}J},P^{I}\}$ depend on spacetime coordinate. Therefore under a local transformation parameterized by these quantities we have

\begin{align}
\partial_{\mu}X^{I} \rightarrow \Lambda^{I}_{\ph{I}J}\partial_{\mu}X^{I} +  \partial_{\mu}\Lambda^{I}_{\ph{I}J}X^{J}+\partial_{\mu}P^{I}
\end{align}
We can introduce the following fields $\{\omega^{I}_{\ph{I}J\mu},\theta^{I}_{\ph{I}\mu}\}$ to define a Poincar\'{e}-covariant derivative:

\begin{align}
D^{({\cal P})}_{\mu}X^{I} &\equiv \partial_{\mu}X^{I} + \omega^{I}_{\ph{I}J\mu}X^{J} + \theta^{I}_{\mu} \label{poin_codiv}
\end{align}
Under a transformation represented by $\{\Lambda^{I}_{J}(x),P^{I}(x)\}$ we have

\begin{align}
D^{({\cal P})}_{\mu}X^{I} &\rightarrow \Lambda^{I}_{\ph{I}J}D^{({\cal P})}_{\mu}X^{J}
\end{align}
if

\begin{align}
\omega^{I}_{\ph{I}J\mu} &\rightarrow \Lambda^{I}_{\ph{I}K}\omega^{K}_{\ph{K}L\mu}(\Lambda^{-1})^{L}_{\ph{L}J} - \partial_{\mu}\Lambda^{I}_{\ph{I}K}(\Lambda^{-1})^{K}_{\ph{K}J} \label{poin_spin_change}\\
\theta^{I}_{\mu} &\rightarrow \Lambda^{I}_{\ph{I}J}\theta^{J}_{\mu} - \partial_{\mu}P^{I} \label{poin_theta_change}
\end{align}
We note that the transformation (\ref{xiglob}) and definitions (\ref{poin_codiv},\ref{poin_spin_change},\ref{poin_theta_change}) follow from the five dimensional matrix representation of the Poincar\'{e} group which we summarize briefly in Appendix \ref{poin_matrix}.

Actions originally possessing the global Poincar\'{e} invariance (\ref{xiglob}) then possess a local Poincar\'{e} invariance if all free `gauge' indices of combinations of covariant derivatives $D^{({\cal P})}_{\mu}X^{I}$ are absorbed by contraction with symbols $\eta_{IJ}$ and $\eps_{IJKL}$ (which are proposed to be invariant under the local Poincar\'{e} transformations) alongside the promotion of partial derivatives of spinors to covariant ones e.g. $\partial_{\mu}\chi \rightarrow D^{(\omega)}_{\mu}\chi$. Note that $\eta_{IJ}X^{I}X^{J}$ is not Poincar\'{e} invariant and so cannot appear in the action. A potential action for gravity is:

\begin{align}
S_{g}[\omega,\theta&,X] = \int c_{IJKL}D^{({\cal P})}X^{I}\wedge D^{({\cal P})}X^{J} \wedge  R^{KL}(\omega)  \nn\\
= \frac{1}{2}&\int d^{4}x \vareps^{\mu\nu\alpha\beta} c_{IJKL}D^{({\cal P})}_{\mu}X^{I}D^{({\cal P})}_{\nu}X^{J}R^{KL}_{\ph{KL}\alpha\beta}(\omega) \label{poinact}
\end{align}
where 

\begin{align}
c_{IJKL} &\equiv \alpha\big(\eps_{IJKL}+2\beta \eta_{I[K}\eta_{L]J}\big) \\
 R^{IJ}_{\ph{IJ}\alpha\beta}(\omega) &=  2\partial_{[\alpha}\omega^{IJ}_{\ph{IJ}\beta]} +2 \omega^{I}_{\ph{I}K[\alpha}\omega^{KJ}_{\ph{KJ}\beta]}
\end{align}
As in the case of (\ref{eincartan}), this action has been chosen as the one of lowest order possible in curvature. Actions of higher order in curvature are additionally consistent with this gauging procedure and may be  considered. Coupling to matter fields $\chi$ is implemented by the promotion $\partial_{\mu}X^{I} \rightarrow D_{\mu}^{({\cal P})}X^{I}$ and the use of the Lorentz covariant derivative $D^{(\omega)}_{\mu} = \partial_{\mu} + \frac{1}{2}\omega_{IJ\mu}S^{IJ}$ acting on spinors:

\begin{align}
S_{m} &=  S_{m}[\chi,D^{({\cal P})}X^{I},\omega^{IJ}(S)]\\
      &= \int {\cal L}_{m}
\end{align}
The equations of motion obtained by varying $S_{g}+S_{m}$ with respect to $\omega,\theta,X$ respectively are:

\begin{align}
 2c_{KLM[I}X_{J]}D^{({\cal P})}X^{K}\wedge R^{LM}&\nn\\
-2c_{KLIJ}D^{(\omega)}\big(\dpo  X^{K}\wedge \dpo X^{L}\big) &\nn\\
+ \frac{\partial {\cal L}_{m}}{\partial \dpo X^{[I}}X_{J]}+ \frac{\partial{\cal L}_{m}}{\partial \omega^{IJ}(S)} &=0 \\
-2c_{IJKL}\dpo X^{J}\wedge R^{KL} + \frac{\partial{\cal L}_{m}}{\partial\dpo X^{I}} &=0 \\
\dpo\big( -2c_{IJKL}\dpo X^{J}\wedge R^{KL} + \frac{\partial{\cal L}_{m}}{\partial\dpo X^{I}}\big) &=0
\end{align}
where, for example, under a small variation of $\omega^{IJ}$ the variation of a Lagrangian four form ${\cal L}$ is $\delta_{\omega}{\cal L} \equiv (\partial {\cal L}/\partial \omega^{IJ})\wedge \delta \omega^{IJ}$.
To make progress, it's useful to conduct a gauge transformation with $P^{I} = -X^{I}$ so that in the new gauge $X^{I}(x)\overset{*}{=}0$ and $\dpo_{\mu} X^{I} \overset{*}{=} \theta^{I}_{\mu}$ and the equations of motion take the form

\begin{align}
-2c_{KLIJ}D^{(\omega)}\big(\theta^{K}\wedge \theta^{L}\big) + \frac{\partial{\cal L}_{m}}{\partial \omega^{IJ}(S)} & \overset{*}{=}0 \\
-2c_{IJKL}\theta^{J}\wedge R^{KL} + \frac{\partial{\cal L}_{m}}{\partial\theta^{I}} & \overset{*}{=}0 \label{poth} \\
\dpo\big( -2c_{IJKL}\theta^{J}\wedge R^{KL} + \frac{\partial{\cal L}_{m}}{\partial\theta^{I}}\big) & \overset{*}{=}0 \label{poxi}
\end{align}
With the identification $\theta^{I}_{\mu} \overset{*}{=} e_{\mu}^{I}$, these are the equations of Einstein-Cartan theory for a general real value of the parameter $\beta$ (and for $\beta = \pm i$) \cite{Immirzi:1996di}. Note that the formal solution to the $X^{I}$ equation (\ref{poxi}) in this gauge is simply given by the $\theta^{I}$ equation of motion (\ref{poth}). This can be stated as the property of General Relativity that - when such a quantity is well-defined -  the total energy-momentum density of gravity and matter is equal to zero.

Coupling to integer spin matter is via the following tensor

\begin{align}
\tilde{g}_{\mu\nu} &\equiv \eta_{IJ}D^{({\cal P})}_{\mu}X^{I}D^{({\cal P})}_{\nu}X^{J} \nn\\
&\overset{*}{=} \eta_{IJ}\theta^{I}_{\mu}\theta^{J}_{\nu}
\end{align}
which is the extension of the quantity $\tilde{\eta}_{\mu\nu}$ to the case where gravitation via the gauging of global Poincar\'{e} symmetry is present.

\section{Gravity via gauging of global translational invariance}
\label{section_gaugingTranslations}

We now consider the case where one gauges only the {\it in}homogeneous piece $P^I$ of the Poincar{\'e} symmetry transformation (\ref{globpoin}). The tetrad field in this case can be identified simply as $e_{\mu}^I = D^{({\cal T})}_{\mu}X^I \equiv \partial_{\mu}X^I + \theta_{\mu}^I$. As we saw was the case for the full Poincar{\'e}
gauge theory in Section \ref{section_gaugingPoincare}, the $X^{I}$ is redundant, at least for classical purposes, in the sense that generally a translational gauge can be found where $X^{I}\overset{*}{=}0$.

In contrast to the polynomial actions of Section \ref{section_gaugingPoincare}, however, we now have 
to resort to very complicated functionals of $e^I_{\mu}$ demanding also the existence of its matrix inverse $e^{\mu}_{I}$, in order to write down
actions which are dynamically equivalent to General Relativity. Having the translation-invariant $e^{I}_{\mu}$, we can then consider its Levi-Civita connection 
$\Gamma^{I}_{\ph{I}J\mu}(e)$ (which, recall, is the solution to the equation $de^{I}+\Gamma^{I}_{\ph{I}J}\wedge e^{J} =0$). One can deduce that the action
\begin{equation} \label{einsteinaction}
S_\parallel[e] = \frac{1}{2}\int\epsilon_{IJKL}e^I\wedge e^J\wedge\Gamma^K{}_M(e)\wedge\Gamma^{LM}(e) 
\end{equation}
realises the dynamical equivalence. In fact the action (\ref{einsteinaction}) in tensor formalism is known as the Einstein action.
Since $\Gamma^{I}_{\ph{I}J}(e)$ is not tensorial, $S_\parallel$ is invariant only up to a boundary term. More properly, it is 
considered as the gauge-fixed action of {\it symmetric} teleparallel gravity \cite{BeltranJimenez:2019esp}.

Now in the space of non-polynomial functionals, there can be many more alternatives. 
It is well-known that now the teleparallel torsion $T^I \equiv D^{({\cal T})} e^I = D^{({\cal T})}D^{({\cal T})}X^I = D^{({\cal T})}(dX^I + \theta^I) = D^{({\cal T})}\theta^I$
is equivalent to the translation gauge field strength. We can consider an action that is quadratic in this gauge field strength.
In terms of the contortion $K^{IJ}$ defined via $K^I{}_J\wedge e^J = T^I$, the action is now
\begin{equation} \label{teleparallel}
S'_\parallel [e] = \frac{1}{2}\int\epsilon_{IJKL}e^I\wedge e^J\wedge K^K{}_M\wedge K^{LM}\,. 
\end{equation}
This action is equivalent to the  {\it metric} teleparallel gravity \cite{Aldrovandi:2013wha,Aldrovandi:2015wfa,Krssak:2018ywd,BeltranJimenez:2019esp}.
Now $S'_\parallel$ is Lorentz-invariant only up to a boundary term, but it can be made invariant by gauging the full Poincar{\'e} symmetry with the
restriction $R^I{}_J = 0$. A problem with this version of the theory is that the connection cannot be consistently minimally coupled to matter \cite{Obukhov:2002tm,Ferraro:2022xwk,BeltranJimenez:2020sih}. 
We may note that both $S_\parallel$ and $S'_\parallel$ are the (tetrad version of the) Einstein-Hilbert action (\ref{einstein-hilbert}), up to the (different) 
boundary terms which both contain second derivatives of $e_{\mu}^I$. The symmetric and metric teleparallel frameworks can be unified in an extension of teleparallel gravity based on a larger general linear gauge symmetry \cite{BeltranJimenez:2019odq}.

Here our main purpose was to clarify the conceptually different approaches to gauging translations. 
The inhomogeneous property of the translation symmetry does not allow formulating a gauge theory with precisely the structure of 
a Yang-Mills theory of a homogeneous symmetry.
In theories based on gauging the global Poincar{\'e} symmetry and the teleparallel special cases such as $S_\parallel$ and $S'_\parallel$ above, 
one introduces a translation gauge potential $\theta_{\mu}^{I}$, but its role is not to provide a covariant  
but an invariant derivative, in contrast to Yang-Mills theory. 

\section{Gravity via gauging of global Lorentz invariance}
\label{section_gaugingLorentz}
We now consider a new possibility: that where only the Lorentz transformation $X^{I} \rightarrow \Lambda^{I}_{\ph{I}J}X^{J}$ in (\ref{globpoin}) is promoted to a local invariance\footnote{The authors of \cite{Aldrovandi:2004uz,Lu:2017wvx} consider this concept but do not present an action that possesses a General-Relativistic limit. Interestingly, the action suggested in \cite{Aldrovandi:2004uz} corresponds to a topological field theory \cite{NikjooZlosnik23}.}. Analogously to the Poincar\'{e} case, the following local Lorentz covariant derivative can be constructed:

\begin{align}
	D^{({\cal A})}_{\mu}X^{I} = \partial_{\mu}X^{I} + {\cal A}^{I}_{\ph{I}J\mu}X^{J}
\end{align}
Under a transformation represented by $\Lambda^{I}_{\ph{I}J}(x)$ we have

\begin{align}
D_{\mu}^{({\cal A})}X^{I} \rightarrow \Lambda^{I}_{\ph{I}J}D_{\mu}^{({\cal A})}X^{J}
\end{align}
if

\begin{align}
{\cal A}^{I}_{\ph{I}J\mu} &\rightarrow \Lambda^{I}_{\ph{I}K}{\cal A}^{K}_{\ph{K}L\mu}(\Lambda^{-1})^{L}_{\ph{L}J} - \partial_{\mu}\Lambda^{I}_{\ph{I}K}(\Lambda^{-1})^{K}_{\ph{K}J}
\end{align}
Additionally, a covariant derivative $\dlo_{\mu}$ acting on Weyl spinors can be defined using ${\cal A}^{I}_{\ph{I}J\mu}$ so that, for example $\dlo_{\mu} \chi^{A'}$ transforms homogeneously under local $SO(1,3)\simeq SL(2,C)$ transformations.  Matter actions originally possessing the global Poincar\'{e} invariance (\ref{xiglob}) then possesses a local Lorentz invariance under the replacement $\partial_{\mu}X^{I} \rightarrow \dlo_{\mu}X^{I}$ and $\partial_{\mu}\Psi \rightarrow \dlo_{\mu}\Psi$ for spinor fields $\Psi$. To complete the picture it is additionally necessary to introduce an action for the gravity itself. A potential action for gravity is:

\begin{align}
S&_{g}[{\cal A},X] =  \int c_{IJKL}\dlo X^{I} \wedge \dlo X^{J} \wedge R^{KL}({\cal A}) \nn\\
&= \frac{1}{2}\int d^{4}x \vareps^{\mu\nu\alpha\beta}c_{IJKL}D^{({\cal A})}_{\mu}X^{I}D^{({\cal A})}_{\nu}X^{J}R^{KL}_{\ph{KL}\alpha\beta}({\cal A}) \label{loract}
\end{align}
where 

\begin{align}
c_{IJKL} = \alpha(\eps_{IJKL} + 2\beta \eta_{I[K}\eta_{L]J})
\end{align}
\begin{align}
R^{IJ}_{\ph{IJ}\alpha\beta}({\cal A}) &=  2\partial_{[\alpha}{\cal A}^{IJ}_{\ph{IJ}\beta]} +2 {\cal A}^{I}_{\ph{I}K[\alpha}{\cal A}^{KJ}_{\ph{KJ}\beta]}
\end{align}
As in the case of (\ref{poinact}), the gauging of a global symmetry of parameterized field theories does not suggest a unique gravitational field, with additional terms higher order in curvature possible. The action (\ref{loract}) may seem like a surprising proposal for a gravitational action given the absence of a gauge potential $\theta^{I}_{\mu}$ which could clearly correspond to $e^{I}_{\mu}$ in a particular gauge, as in the Poincar\'{e} case. Nonetheless, we will see that four dimensional metric structure and gravitational dynamics described by an extension to General Relativity can emerge from a theory whose gravitational fields are $\{{\cal A}^{IJ}_{\mu},X^{I}\}$. Coupling to matter fields $\chi$ is implemented by the promotion $\partial_{\mu}X^{I} \rightarrow \dlo _{\mu}X^{I}$ and the use of the Lorentz covariant derivative $\dlo_{\mu} = \partial_{\mu} + \frac{1}{2}{\cal A}_{IJ\mu}S^{IJ}$ acting on spinors:

\begin{align}
S_{m} &=  S_{m}[\chi^{(m)},\dlo X^{I},{\cal A}^{IJ}(S)]\\
&= \int {\cal L}_{m}
\end{align}
where $\chi^{(m)}$ are matter fields. Unlike in the Poincar\'{e} case, the Lorentz-invariant $X_{I}X^{I}$ is permitted to appear in actions but we do not consider coupling of this quantity to matter, its presence being inconsistent with the procedure of partial gauging of the original global Poincar\'{e} invariance of the non-gravitational theory. The promotion $\partial_{\mu}X^{I} \rightarrow \dlo_{\mu}X^{I}$ in matter actions suggests that the quantity $g_{\mu\nu} \equiv \eta_{IJ}\dlo_{\mu} X^{I}\dlo_{\nu} X^{J}$ will play the role of the spacetime metric tensor. To help show the relation of this model to General Relativity we introduce the auxiliary field $e_{\mu}^{I}$ which is to equal $\dlo_{\mu} X^{I}$ `on shell' and replaces instances of $\dlo_{\mu} X^{I}$ in $S_{g}$ and $S_{m}$, with this equality implemented via the use of a Lagrange multiplier three-form field $\lambda_{I}$:

\begin{align}
S_{\lambda} &= \int \lambda_{I}\wedge \big(\dlo X^{I}-e^{I}\big)
\end{align}
so the total action $S=S_{g}+S_{\lambda}+S_{m}$ takes the form:

\begin{align}
S[e,{\cal A},X,\lambda,\chi^{(m)}] &= \int \bigg[c_{IJKL}e^{I}\wedge e^{J}\wedge R^{KL} \nn\\
&+ \lambda_{I}\wedge\big(\dlo X^{I}-e^{I}\big)\bigg]\nn\\
& +S_{m}[\chi^{(m)},e,{\cal A}] 
\end{align}
The equations of motion obtained by varying $S$ with respect to $e$, ${\cal A}$, $X$, and $\lambda$ are:

\begin{align}
-2c_{IJKL}e^{J}\wedge R^{KL} + \frac{\partial {\cal L}_{m}}{\partial e^{I}}-\lambda_{I} &=0\label{eqmofirst} \\
-\dlo(c_{IJ[MN]}e^{I}\wedge e^{J}) + \frac{\partial{\cal L}_{m}}{\partial {\cal A}^{MN}(S)}+\lambda_{[M}X_{N]} &=0 \label{spinco1} \\
\dlo\lambda_{I} &=0 \\
\dlo X^{I} -e^{I} &=0 \label{eqmolast}
\end{align}
where, for example,  $\delta_{e}{\cal L}_{m}= \frac{\partial {\cal L}_{m}}{\partial e^{I}}\wedge \delta e^{I}$. We now assume that  $X_{I}X^{I} \neq 0 $ over the region of spacetime of interest so that we may define a projector orthogonal to $X^{I}$:

\begin{align}
{\cal P}^{I}_{\ph{I}J} &= \delta^{I}_{\ph{I}J} -\frac{1}{X_{K}X^{K}}X^{I}X_{J}
\end{align}
We therefore have that:

\begin{align}
e^{I}_{\mu} =\frac{1}{X^{2}}{\cal E}_{\mu} X^{I}+E_{\mu}^{I}
\end{align}
where $E^{I}_{\mu}\equiv {\cal P}^{I}_{\ph{I}J}\dlo_{\mu} X^{J}$, hence $X_{I}E_{\mu}^{I} =0$. By the definition $\dlo_{\mu} X^{I}= e_{\mu}^{I}$ we have
${\cal E}_{\mu} = \frac{1}{2}\partial_{\mu}X^{2}$ and hence:

\begin{align}
g_{\mu\nu} &\equiv \eta_{IJ}e^{I}_{\mu}e^{J}_{\nu} = \eta_{IJ}\dlo X^{I} \dlo X^{J} \nn\\
&= \frac{1}{4 X^{2}}\partial_{\mu}X^{2}\partial_{\nu}X^{2}+ E^{I}_{\mu}E_{I\nu} \label{dlogg}
\end{align} 
To cover distinct cases, we can define $X_{I}X^{I} = \xi {\cal X}^{2}$ where $\xi=-1$ if $X_{I}X^{I}$ is timelike and $\xi=1$ if $X_{I}X^{I}$ is spacelike. 
It's useful to clarify the signature of the tensor $h_{\mu\nu} \equiv E^{I}_{\mu}E_{I\nu}$. The quantities $g_{\mu\nu}$ and $h_{\mu\nu}$ are each Lorentz gauge-independent.  For the case $\xi=-1$ and $X^{I}$ is real, we can find a gauge where $X^{I}\overset{*}{=} \sqrt{-X_{J}X^{J}}\delta^{I}_{0}$ where $\eta_{00}=-1$; therefore, as $X_{I}E^{I}=0$, the signature of $h_{\mu\nu}$ is $(0,+,+,+)$ and it can be considered as a spatial metric orthogonal to the timelike vector $\partial^{\mu}X^{2}$. Alternatively, for the case where $\xi=1$ and $X^{I}$ is real, we can find a gauge where $X^{I}\overset{*}{=}\sqrt{X_{J}X^{J}}\delta^{I}_{1}$ where $\eta_{11}$, implying that in this case the signature of $h_{\mu\nu}$ is $(0,-,+,+)$ and it can be considered as a timelike metric orthogonal to the spacelike vector $\partial^{\mu}X^{2}$.

Only for the values $\beta = \pm i$ does a General-Relativistic limit of the theory exist \cite{Zlosnik:2018qvg}. For illustration, the impact of having other values of $\beta$ will be shown in Section \ref{frw} where a general value is considered in the context of Friedmann-Robertson-Walker (FRW) symmetry. We will see that the recovery of a General-Relativistic limit involves the introduction of a \emph{complex} ${\cal A}_{\mu}^{IJ}$. We may additionally allow $X^{I}$ to be complex valued. The symmetry of the action is necessarily then that of the \emph{complexified} Lorentz group $SO(1,3)_{\mathbb{C}}$ (though we will assume that the spacetime manifold is real). For concreteness we will work with the value $\beta=i$. The option $\beta=-i$ yields an identical gravitational theory though differences are possible when coupling of ${\cal A}_{\mu}^{IJ}$ to spinors is considered. With $\beta=i$ we have

\begin{align}
c_{IJKL} &=  \alpha\big(\eps_{IJKL}+2i \eta_{I[K}\eta_{L]J}\big)
\end{align}
The significance of the values $\beta = \pm i$ is as follows: for a differential form $F^{\idxI\idxJ}$ valued in the Lie algebra of $SO(1,3)$ (i.e. $F^{IJ}=-F^{JI}$) one can decompose the form into \emph{self-dual} (+) and \emph{anti self-dual} (-) parts as $F^{\idxI\idxJ} = F^{+\idxI\idxJ} + F^{-\idxI\idxJ}$, $F^{\pm\idxI\idxJ} =\frac{1}{2}(F^{\idxI\idxJ} \mp i\eps^{\idxI\idxJ}_{\ph{\idxI\idxJ}\idxK\idxL}F^{\idxK\idxL}/2)$ where $ \eps^{\idxI\idxJ}_{\ph{\idxI\idxJ}\idxK\idxL}F^{\pm\idxK\idxL} = \pm 2i F^{\pm\idxI\idxJ}$. It follows then that for $\beta=i$:

\begin{align}
c_{IJKL}R^{KL}({\cal A}) = 2\alpha\epsilon_{IJKL}R^{+KL}({\cal A})
\end{align}
Furthermore, it may be verified that

\begin{align}
R^{+\idxI\idxJ}_{\ph{+\idxI\idxJ}}({\cal A}) = R^{\idxI\idxJ}_{\ph{\idxI\idxJ}}({\cal A}^+)
\end{align}
Therefore in this model only the self-dual connection ${\cal A}^{+IJ}_{\mu}$ appears within the curvature. This is as in the case of the Ashtekar formulation of gravity \cite{Ashtekar:1986yd,Jacobson:1988yy}. However, unlike that theory, the action of the current model also contains the anti self-dual connection which appears within the covariant derivative $\dlo_{\mu}X^{I} = \partial_{\mu}X^{I} + {\cal A}^{+IJ}_{\mu}X_{J}+{\cal A}^{-IJ}_{\mu}X_{J}$.

By taking the anti-self dual part of (\ref{spinco1}) and assuming that ${\cal A}^{IJ}_{\mu}(S)= {\cal A}^{+IJ}_{\mu}$ i.e. covariant derivatives of spinor fields are to be built using  ${\cal A}^{+IJ}_{\mu}$ (which is a consistent choice for coupling to spinor fields \cite{Ashtekar:1989ju}) it follows that $\lambda_{I} \propto X_{I}$ and so we may define the field $\varrho$ via $\lambda_{I}= (\varrho/\calx) X_{I}$, following which the equations of motion become:

\begin{align}
-4\alpha \eps_{IJKL}e^{J}\wedge R^{KL}({\cal A}^{+}) + \frac{\partial {\cal L}_{m}}{\partial e^{I}}-\frac{1}{\calx}X_{I}\varrho &=0  \label{eindust} \\
-2\alpha\dlosd(\eps_{IJ[KL]}e^{I}\wedge e^{J})^{+} + \frac{\partial{\cal L}_{m}}{\partial {\cal A}^{+KL}} &=0 \label{spinco}\\
\partial_{\mu}\calx\partial^{\mu}\calx &= \xi \label{calxnorm} \\
d\varrho &=0  \label{drho}\\
E_{I} \wedge \varrho  &=0 \label{wedgenor}
\end{align}
where indices are raised with $g^{\mu\nu}$, taken to be the matrix inverse of (\ref{dlogg}). Equation (\ref{spinco}) in the absence of coupling of the source term due to spinor coupling to the ${\cal A}^{+IJ}_{\mu}$ field\footnote{The inclusion of such sources will modify the solution for ${\cal A}^{+IJ}_{\mu}$ so that a term involving spinor currents will appear when the metric Einstein equations are ultimately recovered.} implies \cite{Romano:1991up} that the solution for the self-dual ${\cal A}^{+IJ}_{\mu}$ is given by the self dual part of the Levi-Civita spin connection $\Gamma^{IJ}_{\mu}(e,\partial e)$ which is defined to be the solution to the equation $de^{I}+\Gamma^{I}_{\ph{I}J}\wedge e^{J}=0$ \cite{Romano:1991up}. We can make contact with standard notation by writing the three-form Einstein equation as a tensor equation:

\begin{align}
4\alpha \eps_{IJKL}\vareps^{\mu\nu\alpha\beta} e^{J}_{\mu}R^{KL}_{\ph{KL}\nu\alpha}({\cal A}^{+}) &= \frac{2}{3!}\bigg[\frac{\partial {\cal L}_{m}}{\partial e^{I}}\bigg]_{\mu\nu\alpha}\vareps^{\mu\nu\alpha\beta}\nn\\
&+\frac{2}{3!}X_{I}\varrho_{\mu\nu\alpha}\vareps^{\mu\nu\alpha\beta} \label{eindev}
\end{align}
We now make the following ansatz for $\varrho$ which satisfies (\ref{wedgenor}):

\begin{align}
\varrho_{\mu\nu\alpha} = -\frac{1}{2}\xi\sqrt{-g}\vareps_{\mu\nu\alpha\beta}\partial^{\beta}\calx \rho
\end{align}
Now, multiplying (\ref{eindev}) by $e_{\zeta}^{I}$ and using $X_{I}e^{I}_{\zeta}= \frac{1}{2}\xi \partial_{\zeta}\calx^{2}$ as well as defining $\alpha \equiv 1/(64\pi G)$ we have:
 
\begin{align}
\bar{R}_{\mu\nu}-\frac{1}{2}\bar{R}g_{\mu\nu} &=8\pi G\big( T^{(m)}_{\mu\nu}+\rho \partial_{\mu}\calx\partial_{\nu}\calx \big) \label{eineqs}
\end{align}
where $\bar{R}^{IJ}_{\ph{IJ}\mu\nu}$  is the curvature two-form associated with $\Gamma^{IJ}_{\mu}(e,\partial e)$, $R^{\mu}_{\ph{\mu}\nu} \equiv R^{\sigma\mu}_{\ph{\sigma\mu}\sigma\nu}$, and we've defined the stress energy tensor for matter fields:

\begin{align}
T^{(m)}_{\mu\nu} &=\frac{1}{3!2\sqrt{-g}}e_{(\nu}^{I}g_{\mu)\sigma}\vareps^{\alpha\beta\gamma\sigma}\bigg[\frac{\partial {\cal L}_{m}}{\partial e^{I}}\bigg]_{\alpha\beta\gamma}
\end{align} 
The equation $d\varrho=0$ becomes:

\begin{align}
0 &=\partial_{\mu}\bigg(\sqrt{-g}\rho \partial^{\mu}\calx\bigg) \label{euler}
\end{align}
We see then from (\ref{eineqs}) and (\ref{euler}) that when $\xi=-1$, Einstein's equations in the presence of an additional dust-like fluid component with density $\rho$ and four-velocity $V_{(\xi=-1)\mu}=\partial_{\mu}\calx$ are recovered. Interestingly, an effective dust-like fluid can emerge in other modifications to Einstein's theory \cite{Mukohyama:2009mz,Chamseddine:2014vna}. Additionally it follows from (\ref{calxnorm}) that 

\begin{align}
V^{\mu}_{(\xi=-1)}\bar{\nabla}_{\mu}V^{\nu}_{(\xi=-1)} = 0
\end{align} 
where $\bar{\nabla}_{\mu}$ is the covariant derivative according to the Christoffel symbols $\Gamma^{\alpha}_{\mu\nu}(g,\partial g)$ i.e. $V^{\mu}_{(\xi=-1)}$ describes timelike geodesic curves in spacetime.
Alternatively, for the case $\xi=1$ ($X_{I}X^{I}>0$), the equations (\ref{eineqs}) and (\ref{euler}) still apply but with a different interpretation: the vector $V^{\mu}_{(\xi=1)}$ is spacelike and, satisfying $V^{\mu}_{(\xi=1)}\bar{\nabla}_{\mu}V^{\nu}_{(\xi=1)}=0$ the fields describe spacelike geodesic curves in spacetime. As such, the source term due to $\rho$ in (\ref{eineqs}) is more readily interpreted as a `dark pressure'. 
We will see in Section \ref{minkowski} that there exist simple solutions where in some parts of spacetime $X^{I}X_{I}<0$, in others $X^{I}X_{I}>0$ and in others $X^{I}X_{I}=0$ (either by $X^{I}$ vanishing or being null) - therefore in such cases the projector ${\cal P}^{IJ}$ cannot be globally defined.

Something that may appear confusing is the role of ${\cal A}_{\mu}^{+IJ}$ and ${\cal A}_{\mu}^{-IJ}$ in producing metric structure (via $\dlo X^{I}$) but also reproducing, via ${\cal A}_{\mu}^{+IJ}$, the self-dual Levi-Civita connection $\Gamma^{+IJ}_{\mu}$, which depends on spacetime derivatives of the same structure. Consider the case where $X_{I}X^{I}<0$. It's helpful to choose a partial gauge fixing where $X^{I} \overset{*}{=} T(x^{\mu})\delta^{I}_{0}$ (where $\eta_{00}=-1$). Then from the definition $e_{\mu}^{I}\equiv \dlo_{\mu}X^{I}$ we have:

\begin{align}
e_{\mu}^{0} &\overset{*}{=}  \partial_{\mu}T \\
e_{\mu}^{i} &\overset{*}{=} {\cal A}^{i}_{\ph{i}0\mu}T
\end{align}
It then follows from the self-dual part of the equation of motion for ${\cal A}^{+IJ}$  that:

\begin{align}
{\cal A}^{+0i}_{\mu} &\overset{*}{=}
\frac{1}{2}\bigg(\Gamma^{0i}_{\ph{0i}\mu}+i\frac{1}{2}\eps^{i}_{\ph{i}jk}\Gamma^{jk}_{\ph{jk}\mu}\bigg)=
\Gamma^{+0i}_{\ph{+0j}\mu}\\
{\cal A}^{+ij}_{\mu} &\overset{*}{=} \frac{1}{2}\bigg(\Gamma^{ij}_{\ph{ij}\mu}-i\eps^{ij}_{\ph{ij}k}\Gamma^{0k}_{\ph{0k}\mu}\bigg)= \Gamma^{+ij}_{\ph{+ij}\mu}
\end{align}
where $\Gamma_{\mu}^{+IJ} = \frac{1}{2}(\Gamma_{\mu}^{IJ} -i \eps^{IJ}_{\ph{IJ}KL}\Gamma_{\mu}^{KL}/2)$. This implies that

\begin{align}
{\cal A}^{IJ} &\overset{*}{=}  \begin{pmatrix} 
0 & \frac{1}{T}e^{i} \\
-\frac{1}{T}e^{i} & \Gamma^{ij} -i\eps^{ij}_{\ph{ij}k}(\Gamma^{0k}-\frac{1}{T}e^{k})\\
\end{pmatrix}
\end{align}
and hence

\begin{align}
{\cal A}^{-0i}_{\mu} &\overset{*}{=} \frac{1}{2}\bigg(\bigg(\frac{2}{T}e^{i}_{\mu} -\Gamma^{0i}_{\mu}\bigg)
-i\frac{1}{2}\eps_{ijk}\Gamma^{jk}_{\mu} \bigg)\\
{\cal A}^{-ij}_{\mu} &\overset{*}{=} \frac{1}{2}\bigg(\Gamma^{ij}_{\mu}   +i\eps^{ij}_{\ph{ij}k}\bigg(\frac{2}{T}e^{k}_{\mu}-\Gamma^{0k}_{\mu}\bigg)\bigg)
\end{align}
By comparison, in the Einstein-Cartan model of gravity based on the action (\ref{eincartan}) in the absence of spinorial sources, the spin connection has the real torsion-free solution $\omega^{IJ}_{\mu} = \Gamma^{IJ}_{\mu}$ and hence $\omega_{\mu}^{-IJ} = [\omega^{+IJ}_{\mu}]^{*}$ (where $*$ denotes complex conjugation). In the self-dual Ashtekar model only the self-dual connection $A^{+IJ}_{\mu}$ appears in the entire formalism and in the absence of spinorial sources the theory's equations of motion imply $A^{+IJ}_{\mu}=\Gamma^{+IJ}_{\mu}$ \footnote{Interestingly,
an approach based on an $SO(1,3)$ connection ${\cal A}^{IJ}_{\mu}$, a field $X^{I}$, and an \emph{independent} field $e^{I}_{\mu}$ yields a variety of novel phenomenology \cite{Klinkhamer:2018mmk}.}.

\section{An Example: Friedmann-Robertson-Walker symmetry}
\label{frw}
By way of illustration, we can consider the equations of this theory in a situation of high spacetime symmetry. Consider a case where $\eta_{IJ}X^{I}X^{J} < 0$ throughout spacetime, in which case one can globally find the Lorentz gauge $X^{I} \overset{*}{=} T(x) \delta^{I}_{0}$. We adopt the following ansatz for the gravitational fields:
\begin{align}
T &= T(t)\\
{\cal A}^{0i}&=bE^i \\
{\cal A}^{12}&=-\frac{K(r)}{r}E^2-cE^3 \\
{\cal A}^{13}&=-\frac{K(r)}{r}E^3+cE^2\\ {\cal A}^{23}&=-\frac{\cot\theta}{r}E^3-cE^1
\end{align}
where $b=b(t)$, $c=c(t)$, and we've defined the following comoving spatial coordinate basis one-forms:
\begin{align}
E^0=dt\quad E^1=\frac{dr}{K(r)}\quad E^2=rd\theta\quad E^3=r\sin\theta d\varphi.
\end{align}
where $K(r)=\sqrt{1-kr^2}$ with $k$ being the constant of spatial curvature. This corresponds to restriction to Friedmann-Robertson-Walker symmetry. The curvature two-form $R^{IJ}({\cal A})$ becomes:

\begin{align}
R^{0i}&=\dot bE^0E^i+bc\epsilon^i_{\ph ijk}E^jE^k\\ R^{ij}&=(k-c^2+b^2)E^iE^j-\dot c \epsilon^{ij}_{\ph{ij}k}E^0E^k
\end{align}
Therefore we can identify the spacetime metric as

\begin{align}
g_{\mu\nu} &\equiv \eta_{IJ}\dlo_{\mu} X^{I}\dlo_{\nu} X^{J} \nn\\
&=- (\dot{T})^{2}\partial_{\mu}t \partial_{\nu}t + \delta_{ij} (bT)^{2} E^{i}_{\mu}E^{j}_{\nu} \label{cosmometric}
\end{align}
where $\bar{h}_{\mu\nu} \equiv \delta_{ij}E^{i}_{\mu}E^{j}_{\nu}$ is the (constant in time) metric of either flat three dimensional Euclidean space, the three-sphere, or the three-hyperboloid for $k=0$, $k>0$, and $k<0$ respectively. The form of (\ref{cosmometric}) suggests that we should look to consider the combination $a \equiv bT$ to be equal to the metric scale factor. The action for gravity is, considering a general value of $\gamma \equiv 1/\beta$:

\begin{widetext}
\begin{align}
S_{g} &=\frac{1}{32\pi G}\int \bigg(\eps_{IJKL}+\frac{2}{\gamma}\eta_{IK}\eta_{JL}\bigg)\dlo X^{I}\wedge \dlo X^{J} \wedge R^{KL} \\
&=\frac{1}{16\pi G} \int \bigg(T\dot{T}b\bigg(k-c^{2}+b^{2}-\frac{2}{\gamma}cb\bigg) +b^{2}T^{2}\bigg(\dot{b}  - \frac{1}{\gamma}\dot c\bigg)\bigg) \eps_{ijk}dt \wedge E^{i}\wedge E^{j}\wedge E^{k}
\end{align}
\end{widetext}
For illustrative purposes consider matter to be described by a scalar field with Lagrangian density

\begin{align}
L_{\phi} &=  \sqrt{-g}F(\phi,{\cal K}) \label{scalarlag}
\end{align}
where ${\cal K} \equiv -g^{\mu\nu}\partial_{\mu}\phi\partial_{\nu}\phi$. In FRW symmetry we have $\phi=\phi(t)$ so ${\cal K} = \dot{\phi}^{2}/\dot{T}^{2}$ and $\sqrt{-g} = \dot{T}a^{3}$ so

\begin{align}
L_{\phi} &=  \dot{T}a^{3}F(\phi,(\dot{\phi}/\dot{T}))
\end{align}
Putting everything together we have, up to a boundary term:

\begin{widetext}
\begin{align}
S_{g}+S_{\phi} 
&\overset{b}{=}  \frac{6}{32\pi G}\int  \sqrt{\bar{h}}d^{3}x dt\bigg[2\dot{T}a\bigg(k-c^{2}+\bigg(\frac{a}{T}\bigg)^{2}-\frac{2}{\gamma}c\frac{a}{T}\bigg) +\bigg(2a^{2}\frac{d}{dt}\bigg(\frac{a}{T}\bigg)  + \frac{2}{\gamma}\frac{da^{2}}{dt}c\bigg)\bigg] \nn\\
& + \int \sqrt{\bar{h}} d^{3}xdt \dot{T}a^{3}F(\phi,(\dot{\phi}/\dot{T})) 
\end{align}
\end{widetext}
where $\overset{b}{=}$ means equal to up to a boundary term and $\bar{h}= \mathrm{det}(\bar{h}_{\mu\nu})$. Varying with respect to $c$ we have 

\begin{align}
b+ \gamma c= \dot{a}/\dot{T} \label{spincofrw}
\end{align} 
This is the restriction of the equation of motion (\ref{spinco}) to FRW symmetry and illustrates how this combination of parts of ${\cal A}^{IJ}$ are solvable in terms of derivatives of the metric tensor. In this case we may use (\ref{spincofrw}) to solve for the field $c(t)$ algebraically and eliminate it from the action, which then after integration by parts takes the form

\begin{align}
S_{g}+S_{\phi} &
\overset{b}{=}\int \sqrt{\bar{h}}d^{3}xdt a^{3}\dot{T} \bigg[\frac{3}{8\pi G}\bigg(\frac{1}{\gamma^{2}}\frac{1}{a^{2}}\frac{\dot{a}^{2}}{\dot{T^{2}}}+\frac{k}{a^{2}}\bigg)\nn\\
&+\frac{(1+\gamma^{2})}{8\pi G\gamma^{2}}\frac{1}{T^{2}}+F\bigg(\phi,\frac{\dot{\phi}}{\dot{T}}\bigg)\bigg]
\end{align}
The recovery of the FRW-symmetric action for General Relativity is only recovered for $\gamma = \pm i$, which we henceforth adopt, yielding:

\begin{align}
S_{g}+S_{\phi} &\overset{b}{=}\int \sqrt{\bar{h}}d^{3}xdt a^{3}\dot{T} \bigg[\frac{3}{8\pi G}\bigg(-\frac{1}{a^{2}}\frac{\dot{a}^{2}}{\dot{T^{2}}}+\frac{k}{a^{2}}\bigg)\nn\\
&+F\bigg(\phi,\frac{\dot{\phi}}{\dot{T}}\bigg)\bigg]  \label{frwppn}
\end{align}
Note that the action (\ref{frwppn}) has reduced to that of parameterized particle mechanics with $t$ playing the role of the parameter $\lambda$. We can introduce new fields $P_{T}$ and $N$ such that $P_{T}$ is a Lagrange multiplier term enforcing the definition of the `time velocity' $N=\dot{T}$, with the action becoming:

\begin{align}
S[P_{T},T,N,a,\phi] &\overset{b}{=}\int  \sqrt{\bar{h}}d^{3}xdt  \bigg[P_{T}\bigg(\dot{T}-N\bigg)\nn\\
& + a^{3}N \bigg(\frac{3}{8\pi G}\bigg(-\frac{1}{N^{2}}\frac{1}{a^{2}}\dot{a}^{2}+\frac{k}{a^{2}}\bigg)\nn\\
&+F\bigg(\phi,\frac{\dot{\phi}}{N}\bigg)\bigg)\bigg]
\end{align}
Varying $T$ and $N$ we have:
\begin{align}
\frac{3}{8\pi G N^{2}}\bigg(\frac{\dot{a}^{2}}{a^{2}}+\frac{k}{a^{2}}\bigg) &=   \bigg(\frac{\dot{\phi}}{N}\frac{\partial F}{\partial (\dot{\phi}/N)}-F\bigg)+\frac{P_{T}}{a^{3}} \label{darkfriedmann} \\
\dot{P}_{T} &=0  \label{dotp}
\end{align}
whilst equations of motion obtained by varying $a$ and $\phi$ respectively are identical to those in General Relativity coupled to a scalar field with Lagrangian density (\ref{scalarlag}). Equations (\ref{darkfriedmann}) and (\ref{dotp}) arise from the field equations (\ref{eindust}) and (\ref{drho}) restricted to FRW symmetry. Additionally, $P_{T}$ itself is the analogue  of the integration constant $E$ in parameterized particle mechanics and it has an observational effect: it would be interpreted as a dark matter component in the universe.

\section{Minkowski Solutions}
\label{minkowski}
We now show that the model based on local Lorentz symmetry admits several distinct field configurations that solve the field equations and result in the spacetime metric being that of Minkowski space.  Recall that in the case of the non-gravitational parameterized field theory, a Minkowski metric was recovered via

\begin{align}
\tilde{\eta}_{\mu\nu} = \eta_{IJ}\frac{\partial X^{I}}{\partial x^{\mu}}\frac{\partial X^{J}}{\partial x^{\nu}}
\end{align}
We will now show that two distinct solutions in the gravitational theory lead to the recovery of Minkowski space. To aid visualization, figures showing the profile of the field $X^{I}$ in Minkowski space are included in Appendix \ref{thefigure}.

\subsection{Vanishing curvature}
Clearly a solution to the vacuum equations of motion (\ref{eqmofirst})-(\ref{eqmolast}) for $\beta=\pm i$ is if the curvature two-form $R^{I}_{\ph{I}J}=0$. Then, one can find a gauge where ${\cal A}^{IJ}_{\mu} \overset{*}{=} 0$ and in this gauge

\begin{align}
g_{\mu\nu} &\overset{*}{=} \eta_{IJ}\partial_{\mu}X^{I} \partial_{\mu}X^{J}
\end{align}
The equations of motion admit solutions where $X^{I}$ can coordinatize the entire spacetime such that $g_{\mu\nu}= \eta_{\mu\nu}$. Note that here $X^{I}=0$ at a single point in spacetime. Such a solution is not unique. Additional $X^{I}$ related to the original solution by $X^{I}\rightarrow \Lambda^{I}_{\ph{J}J}X^{J}+P^{I}$ - where $\Lambda^{I}_{\ph{I}J}\in SO(1,3)_{\mathbb{C}},P^{I}\in \mathbb{C}^{4}$ and $\partial_{\mu}\Lambda^{I}_{\ph{I}J} = \partial_{\mu}P^{I}=0$ - are also solutions i.e. there are a family of solutions related by global complexified Poincar\'{e} transformations.

\subsection{Non-vanishing curvature}
\label{curvedflat}
An alternative possibility is to consider the case of timelike $X^{I}$ in FRW symmetry and obtain a solution where the scale factor $a(t) = \mathrm{Cst.}$, implying that the metric tensor $g_{\mu\nu}$ takes Minkowski form. Indeed it can readily be seen that equations (\ref{darkfriedmann}) and (\ref{dotp}) for the case of no matter sources (e.g. in the case of the scalar field of Section \ref{frw} this would mean $F=0$) and for zero spatial curvature they possess a solution $P_{T} =0$.  Then from the definition of $b$ and the equation of motion for the field $c$ we have:

\begin{align}
b &= \frac{1}{t}, \quad c = -\frac{i}{t}
\end{align}
Due to the vanishing spatial curvature, we can pick Cartesian spatial coordinates $\{x^{i}\}$ such that $E^{i}=dx^{i}$ and the curvature two-form takes the form:
\begin{align}
R^{0i}&= -\frac{1}{t^{2}}dt\wedge dx^{i}-\frac{i}{t^{2}}\epsilon^i_{\ph ijk}dx^{j}\wedge dx^{k} \label{roimin}\\ 
R^{ij}&=\frac{2}{t^{2}}dx^{i}\wedge dx^{j}-\frac{i}{t^{2}} \epsilon^{ij}_{\ph{ij}k}dt\wedge  dx^{k}\label{rijmin}
\end{align}
Here $X^{I}$ now vanishes on the 3-surface $t=0$. Remarkably, the curvature tensor is non-zero for this solution with flat spacetime metric\footnote{This is the opposite of the case of teleparallel gravity (discussed in more detail in Section \ref{section_gaugingTranslations}) where the spacetime curvature is zero but nonetheless metrics with non-vanishing Riemannian curvature (i.e. curvature built from the Christoffel symbols) exist as solutions to the field equations \cite{Aldrovandi:2013wha,Aldrovandi:2015wfa,Krssak:2018ywd,BeltranJimenez:2019esp}.}. Note that the curvature diverges as $t\rightarrow 0$. However, nonetheless the gravitational action $S_{g}$ remains zero for all moments of time as it only depends on the \emph{self-dual} curvature which - as can be verified from (\ref{roimin}) and (\ref{rijmin}) - always vanishes. The existence of solutions with a maximally symmetric metric in the presence of fields
which spontaneously break local Lorentz invariance is reminiscent of ghost condensate \cite{Arkani-Hamed:2003pdi} and Einstein-Aether \cite{Jacobson:2000xp,Foster:2005dk} models which permit Minkowski space as a solution to the field equations despite the presence of, respectively, a scalar field $\phi$ with non-zero time derivative or vector field $A^{\mu}$ with timelike expectation value.
\subsection{Perturbations around Minkowski background solutions}
We now consider small perturbations to the Minkowski background solutions.
In regions where the background solution $X^{I}= \bar{X}^{I}$ satisfies $\bar{X}_{I}\bar{X}^{I} \neq 0$ the equation (\ref{euler}) perturbed to linear order in $\rho$ and $X^{I}$ reads:

\begin{align}
\partial_{\bar{\calx}}\delta\rho=0
\end{align}
where recall that by definition $\bar{X}_{I}\bar{X}^{I}= \xi \bar{\cal X}^{2}$ where $\xi=-1$ if $\bar{X}_{I}\bar{X}^{I}$ is timelike and $\xi=1$ if $\bar{X}_{I}\bar{X}^{I}$ is spacelike. Consider the vanishing-curvature solution. For, say, the `upper' region in the solution where $\bar{X}_{I}\bar{X}^{I}<0$ ($t>r$, $t>0$ in spherical coordinates $(t,r,\theta,\phi$)), we can coordinatize spacetime by coordinates $(\bar{\calx},\chi,\theta,\phi)$ such that the Minkowski metric in these coordinates is:

\begin{align}
\eta  &=  -d\bar{\calx}^{2} + \bar{\calx}^{2}(d\chi^{2}+\sinh^{2}\chi d\Omega^{2})\\
   &=  -dt^{2}+ dr^{2}+r^{2}d\Omega^{2}
\end{align}
So

\begin{align}
r  &=  \bar{\calx}\sinh\chi\\
t &=  \bar{\calx}\cosh\chi
\end{align}
where $
\bar{\calx} = (t^{2}-r^{2})^{1/2}$ ($0<\bar{\calx} < \infty$), and $\chi = \cosh^{-1}\big(t/(t^{2}-r^{2})^{1/2}\big)$  $(-\infty < \chi < \infty)$.
So the perturbed stress energy tensor in the $(\bar{\calx},\chi)$ coordinate basis has one non-vanishing component  ${\cal T}_{\bar{\calx}\bar{\calx}} =\delta \rho$ which in the inertial coordinate basis yields:

\begin{align}
{\cal T}_{tt} &=  \frac{t^{2}}{t^{2}-r^{2}}\delta\rho \\
{\cal T}_{rr} &=  \frac{r^{2}}{t^{2}-r^{2}}\delta\rho
\end{align}
where $\delta\rho = \delta\rho(\chi,\theta,\phi)$ and with all other components of ${\cal T}_{\mu\nu}$ vanishing.

In the region for the vanishing-curvature solution solution where $\bar{X}_{I}\bar{X}^{I}>0$, we can coordinatize spacetime by coordinates $(\bar{\calx} = \sqrt{r^{2}-t^{2}},\zeta,\theta,\phi)$ (where $r>0$ and $r>t$ in spherical coordinates $(t,r,\theta,\phi)$) such that the Minkowski metric in these coordinates is:

\begin{align}
\eta  &=  d\bar{\calx}^{2} + \bar{\calx}^{2}(-d\zeta^{2}+\cosh^{2}\zeta d\Omega^{2})\\
&=  -dt^{2}+ dr^{2}+r^{2}d\Omega^{2}
\end{align}
where

\begin{align}
r &= \bar{\calx} \cosh\zeta \\
t &= \bar{\calx} \sinh\zeta
\end{align}
So $\bar{\calx} = \sqrt{r^{2}-t^{2}}$ ($0<\bar{\calx} < \infty$) and $\zeta = \cosh^{-1}(r/(r^{2}-t^{2})^{1/2})$ ($-\infty < \zeta < \infty$). Now, the equations of motion imply that $\delta\rho$ is constant along the spacelike vector $\partial^{\mu}\bar{\calx}$ so the perturbed stress energy tensor in the $(\bar{\calx},\zeta)$ basis has one non-vanishing component ${\cal T}_{\bar{\calx}\bar{\calx}} = \delta\rho$, hence in this region

\begin{align}
{\cal T}_{tt} &= \frac{t^{2}}{r^{2}-t^{2}}\delta\rho\\
{\cal T}_{rr} &=  \frac{r^{2}}{r^{2}-t^{2}}\delta\rho
\end{align}
where $\delta\rho = \delta\rho(\zeta,\theta,\phi)$ and with all other components of ${\cal T}_{\mu\nu}$ vanishing. To stop divergence of ${\cal T}_{\mu\nu}$ there should be  appropriate fall off of $\delta\rho$ to zero as the limiting values of $\chi$ and $\zeta$ respectively are approached to compensate for the accompanying diverging $1/|r^{2}-t^{2}|$. It seems reasonable to require that ${\cal T}_{\mu\nu}^{(\bar{X}^{2}>0)}$ joins smoothly with ${\cal T}_{\mu\nu}^{(\bar{X}^{2}<0)}$ across the $\bar{X}^{2}=0$ null surface and at $\bar{X}^{I}=0$  and for the resultant global ${\cal T}_{\mu\nu}$ to source a smooth metric perturbation $\delta g_{\mu\nu}$ as solutions to Einstein's equations (which apply in each region where $\bar{X}^{2}\neq 0$).

The case of non-vanishing curvature is more straightforward. As now the background $\bar{\calx}$ coordinate can simply be identified with $t$ and as such we have:

\begin{align}
{\cal T}_{tt} &=  \delta \rho
\end{align}
with all other components of ${\cal T}_{\mu\nu}$ vanishing and, to linear order in perturbations $\partial_{t}\delta\rho =0$.

\section{Other solutions}
\label{other}
It is also possible to find solutions to the field equations where the four dimensional metric corresponds to that of flat four dimensional \emph{Euclidean} space i.e. where coordinates exist so that the metric can globally be put in the form $g_{\mu\nu}=\eta_{IJ}\dlo_{\mu} X^{I}\dlo_{\nu} X^{J} = \mathrm{diag}(1,1,1,1)$. This may be recovered from the non-vanishing curvature Minkowski solution via $(T,b) \rightarrow (iT,\pm ib)$ or considering a zero-curvature solution for which in the gauge ${\cal A}^{IJ}_{\mu} \overset{*}{=}0$ we have $X^{I}=(it,x^{i})$ where $(t,x^{i})$ comprise a set of inertial coordinates in spacetime. Interestingly the field $X^{I}$ in the non-zero curvature solution introduces a `preferred' (imaginary) time coordinate but nonetheless the resultant Euclidean geometry with four dimensional metric $\delta_{\mu\nu}$ possesses symmetry under the group of diffeomorphisms generating global $ISO(4)$ coordinate transformations.

In a more general context, recall that the spacetime metric tensor takes the following form when $X^{2}\equiv X_{I}X^{I} \neq 0$.

\begin{align}
g_{\mu\nu} &= \frac{1}{4 X^{2}}\partial_{\mu}X^{2}\partial_{\nu}X^{2}+ E^{I}_{\mu}E_{I\nu}
\end{align} 
Consider the case where there exists an $SO(1,3)_{C}$ gauge where $X^{I}\equiv i S(x^{\mu})\delta^{I}_{0}$ where $S(x^{\mu})$ is assumed real and hence $X_{I}X^{I} = S^{2}$. In this gauge we have

\begin{align}
E^{I}_{\mu} &\overset{*}{=}  iSA^{I}_{\ph{I}0\mu}
\end{align}
Therefore the only non-vanishing $E^{I}_{\mu}$ are $E^{i}_{\mu}$ where $i,j,k = 1\dots 3$ and  

\begin{align}
g_{\mu\nu} = \frac{1}{S^{2}}\partial_{\mu}S^{2}\partial_{\nu}S^{2}- S^{2}\eta_{ij}A^{i}_{\ph{i}0\mu}A^{j}_{\ph{i}0\nu}
\end{align}
So if $A^{i}_{\ph{i}0\mu}$ in this gauge are purely imaginary then the spacetime metric is real and of Euclidean signature.
\section{Phenomenology}
\label{section_phenomenology}
From equations (\ref{eineqs}) and (\ref{euler}) we see that in the regime $X_{I}X^{I}<0$, the extension to General Relativity arrived at via the action (\ref{loract}) when $\beta=\pm i$  is the presence of a pressureless, perfect fluid source in Einstein's equations. The equations of motion for this system equivalently follow from the action

\begin{align}
S[g,\rho,{\cal X}] &= \int d^{4}x \sqrt{-g}\bigg[\frac{1}{16\pi G}R - \rho(\partial_{\mu}{\cal X}\partial^{\mu}{\cal X}+1)\bigg] \label{grappf}
\end{align}
We now consider the phenomenology associated with the action (\ref{grappf}). Notably, the standard model of cosmology consists of General Relativity, the fields of the standard model of particle physics, a positive cosmological constant $\Lambda$, and new degrees of freedom which behave precisely as a pressureless, perfect fluid on cosmological scales (dark matter) \cite{Planck:2018vyg}. In General Relativity, the introduction of a non-zero cosmological constant is an `economical' explanation for data suggesting late-time acceleration of the universe because it requires no new degrees of freedom beyond those present in General Relativity to be introduced. Analogously, the dark matter effect appearing from the action (\ref{loract}) is to be considered an inherent part of gravitation: it arises from the gravitational degrees of freedom $\{\omega^{IJ},X^{I}\}$ which define the geometry itself.
Can the dark matter effect arising from  (\ref{loract}) (and hence from (\ref{grappf})) be considered a realistic dark matter candidate? 

A good approximation to all dark matter models on large, cosmological scales is expected to be the hydrodynamical description in which the dark matter is described by a fluid with density $\rho(x)$ and four velocity $u^{\mu}$ where the four velocity  obeys the geodesic equation according to the metric $g_{\mu\nu}$.  Generally a configuration $u^{\mu}$ specified on an initial Cauchy surface will evolve so that $\nabla_{\mu}u^{\mu}$ diverges in finite time (the formation of caustics), preventing further evolution of the field via the equation of motion (\ref{euler}) \cite{Hawking:1973uf}. It is not difficult to find initial data so that the pathological behaviour arises on timescales orders of magnitude shorter than the age of the universe \cite{Babichev:2017lrx}  and so the viability of the classical equations of motion following from (\ref{grappf}) is called into question. A possibility is that a \emph{cosmic skeleton} of singular structures would appear in such a scenario; a consequence of this scenario would be that supermassive black holes would form with such ease that the observed mass of the presumed black hole in the centre of the Milky Way galaxies constrains the cosmic abundance of such `irrotational' dark matter to be a small fraction of the total amount in our universe \cite{Sawicki:2013wja}.

It seems likely then that new physics beyond the classical equations of motion (\ref{eineqs}) and (\ref{euler}) must come into play. In particle models of dark matter, the would-be appearance of caustics signifies the breakdown of the hydrodynamical approximation in favour of a description in terms of particles, which may collide or pass through one another. In models where the four velocity is the gradient of a field (e.g. $u_{\mu}=\partial_{\mu}{\cal X}$ in the case of the action (\ref{grappf})) which is to be regarded as `fundamental', the behaviour of the field must depart from that dictated by solutions to the equations (\ref{eineqs}) and (\ref{euler}). 

One possibility is that quantum corrections to the classical equations of motion prevent the formation of caustics. By way of example, one approach \cite{Brown:1994py,Husain:2011tk} has been to construct the canonical formulation of the action (\ref{grappf}) and then implement a time gauge fixing constraint ${\cal X}\overset{*}{=} t$ (which is analogous to the gauge $T \overset{*}{=}\lambda$ in parameterized particle mechanics example given in Section \ref{section_parameterization}) prior to quantization. Hence, if ${\cal X}$ plays the role of time in the putative quantum theory of gravity (and allowed to flow eternally without obstruction), it is not clear that the caustic pathologies which prevent the use of ${\cal X}$ as a global clock in the classical theory can emerge as a limit of the quantum theory. Indeed there is evidence that caustics are indeed avoided when spherical collapse of the pressureless perfect fluid is considered for the model (\ref{grappf}) quantized in accordance with \cite{Husain:2021ojz}. Additionally, it is known in the context of Friedmann-Robertson-Walker symmetry, the big bang singularity associated with the classical equations (\ref{eineqs}) and (\ref{euler}) may be avoided in the quantum theory restricted to this symmetry \cite{Gielen:2020abd,Gielen:2022ouk}. A criticism of such approaches \cite{Isham:1992ms} has been that it has not been clear how degrees of freedom $(\rho,{\cal X})$ could appear in a physical theory and we regard it as encouraging that they arise naturally from a theory based on the action (\ref{loract}).

However, it should be emphasized that this interpretation of (\ref{grappf}) is not universal. Rather, \cite{Blas:2009yd} considered the fluid part of the action decoupled from gravity and constructed the canonical formulation of this part in isolation, recovering a Hamiltonian density ${\cal H} = \Pi_{({\cal X})}\sqrt{1+\partial^{i}{\cal X}\partial_{i}{\cal X}}$, where $\Pi_{({\cal X})}$ is the canonical momentum of ${\cal X}$ and $i$ denotes a spatial coordinate index which is raised with flat Euclidean inverse metric. The authors then consider an expansion around a background solution ${\cal X}=t$, $\Pi_{({\cal X})}=\rho_{0}$ ($\partial_{\mu}\rho_{0}=0$) with $\delta{\cal X}= \chi/\sqrt{\rho_{0}}$, $\delta\Pi_{({\cal X})}= \Pi_{\chi}\sqrt{\rho_{0}}$ where $\rho_{0}$ is to be interpreted as the background density of the pressureless perfect fluid. It follows then that ${\cal H} = \frac{1}{2}\partial^{i}\chi \partial_{i}\chi +\frac{1}{2\sqrt{\rho_{0}}}\Pi_{\chi}\partial^{i}\chi \partial_{i}\chi + \dots$, which suggests that the perturbative expansion breaks down for energy scales $\Lambda \sim \rho_{0}^{1/4}$ which for the current cosmic dark matter density corresponds to $\Lambda \sim 10^{-3}eV$ suggesting that perturbative quantization of the fluid part of (\ref{grappf}) is limited to energy scales $E\ll \Lambda$, which has been argued to be unacceptable for a component of a candidate theory of quantum gravity. It is unlikely that a quantum theory based on this perturbative approach is equivalent to the one based on gauge fixing ${\cal X}=t$ prior to quantization.

Another possibility is that new degrees of freedom beyond those present in (\ref{grappf}) become active in regimes close to the formation of caustics, in effect causing the velocity field $u_{\mu}$ to depart from geodesic motion and leading to caustic avoidance. A well-known example of this is the `UV completion' of (\ref{grappf}) in terms of a massive, complex scalar field $\Phi = \lambda e^{i\phi}$. In curved spacetime the Lagrangian for such a field is 

\begin{align}
L_{\Phi} &= \frac{1}{2}\sqrt{-g}(-g^{\mu\nu}\partial_{\mu}\Phi^{*}\partial_{\nu}\Phi - M^{2}|\Phi|^{2})\nn\\
& = \frac{1}{2}\sqrt{-g}\bigg(-\frac{g^{\mu\nu}\partial_{\mu}\tilde{\lambda} \partial_{\nu}\tilde{\lambda}}{M^{2}} - 
\tilde{\lambda}^{2}\big(g^{\mu\nu}\partial_{\mu}\tilde{\phi} \partial_{\nu}\tilde{\phi} + 1\big)\bigg)
\end{align}
where $\tilde{\lambda} = M\lambda$, $\tilde{\phi}=\phi/M$. In the limit $M\rightarrow \infty$ and with the identification $\tilde{\lambda}^{2} = 2\rho$, $\tilde{\phi}= {\cal X}$, we see that $L_{\Phi}$ tends to the form of the fluid part of $(\ref{grappf})$ and indeed it can be shown that solutions for gravity coupled to $L_{\Phi}$ can approach those of (\ref{grappf}) for sufficiently large $M$. For finite $M$ it follows from the $\tilde{\lambda}$ equation of motion that the `four-velocity' $u_{\mu} \equiv \partial_{\mu}\tilde{\phi}/\sqrt{-(\partial_{\nu}\tilde{\phi}\partial^{\nu}\tilde{\phi})}$ does not satisfy the geodesic equation and it can be shown that caustics associated with this field do not form \cite{Babichev:2017lrx}. 

Thus an alternative to important quantum corrections to (\ref{grappf}) arising would be such a `UV completion' of the model (\ref{loract}) so as to introduce new degrees of freedom to ameliorate the problem of caustics. Such a scenario is not inconceivable: for example, despite the great success of General Relativity, a leading candidate for cosmic inflation and the origin of structure in the universe is the Starobinsky model of inflation \cite{Starobinsky:1980te} which considers a correction $\sqrt{-g}R^{2}$ to the Einstein-Hilbert Lagrangian; this model is equivalent to a scalar tensor theory and the new scalar degree of freedom in gravitation can be of great importance at high energy scales - for example in sourcing large scale structure in the universe \cite{Mukhanov:2005sc}.

A final possibility is that the constraint $u_{\mu}u^{\mu}+1=0$ with $u_{\mu}=\partial_{\mu}{\cal X}$ remains in place so that $u_{\mu}$ always satisfies the geodesic equation but that additional terms in the action become important close to caustic formation so as to create a repulsive gravity effect, stopping $\nabla_{\mu}u^{\mu}$ from diverging. Indeed, a dark matter effect with a number of similar characteristics to that following from (\ref{loract}) was discovered in the context of the projectable Ho\v{r}ava-Lifshitz gravity \cite{Mukohyama:2009mz,Mukohyama:2009tp} where the four velocity of the dark matter fluid takes the form $u_{\mu} = -\partial_{\mu}T$ where $T(x)$ is a scalar field which acts as a preferred time coordinate in spacetime. It has been argued that caustics should be expected to not form in such theories due to a) corrections to the Lagrangian that depend on the extrinsic curvature of surfaces of constant $T$ (and so may include $\nabla_{\mu}u^{\mu}$) which modify classical gravitational dynamics so as to provide a repulsive effect preventing the divergence of $\nabla_{\mu}u^{\mu}$ and b) quantum behaviour of the gravitational degrees of freedom, akin to how the big bang singularity may be avoided in minisuperspace quantum cosmological models of a system comprising General Relativity and dust. As we have discussed, behaviour b) may also arise from the action (\ref{loract}) whilst corrections of the type a) are conceivable: it may be checked that equations of motion for the model $\beta=\pm i$ (\ref{loract}) imply that the extrinsic curvature of surfaces of constant ${\cal X}$ is contained within the torsion $D^{(\cal A)}e^{I} = R^{I}_{\ph{I}J}({\cal A})X^{J}$ and so additional terms in the action of higher order in these parts of the curvature may be able to dynamically prevent singular behaviour in this extrinsic curvature.

It is beyond the current scope of this work to provide a definitive resolution to the question of the corrections that should be expected to equations (\ref{eineqs}) and (\ref{euler}) and how they affect the viability of a dark matter candidate arising from a description of gravity in terms of a spontaneously-broken gauge theory of the Lorentz group. 
The scenario that geodesic motion is modified by repulsive gravity effects in the vicinity of would-be caustics is perhaps most immediately testable given the effect such a modification would have on the propagation of light, leading to a potential gravitational lensing signature.

Finally we comment on possible experimental signatures associated with the Minkowski solution possessing non-zero gauge field curvature $R^{IJ}({\cal A})$ presented in Section \ref{curvedflat}. In a Minkowski coordinate basis $(t,x^{i})$, the solution implies that ${\cal A}^{+IJ}=0$, ${\cal A}^{-IJ}=  (2/t)(n^{[I}E^{J]}+\frac{i}{2}\eps^{IJKL}n_{K}E_{L})$ where $n^{I} = X^{I}/\sqrt{-X_{J}X^{J}}$ where $E_{L}$ are spatial coordinate basis one-forms satisfying $E_{L}n^{L}=0$ whilst the metric $g_{\mu\nu} = \dlo_{\mu} X^{I}\dlo_{\nu} X_{I}= \eta_{\mu\nu}$. Any field that couples to ${\cal A}^{-IJ}$ in isolation (i.e. aside from the coupling to ${\cal A}^{-IJ}$ contained within $g_{\mu\nu}$) will be affected by the background curvature. Here appears an apparent choice in the coupling between spinor fields and gravity. Consider the kinetic term for a $-$ (minus) chirality spinor $\chi^{A'}$. There are two independent possibilities:

\begin{align}
i\eps_{IJKL}e^{J}\wedge e^{K}\wedge e^{L}\wedge\big(\chi^{*A}\sigma^{I}_{AA'}D^{({\cal A}^{-})}\chi^{A'}\big) \label{tospinors1}
\end{align}
\begin{align}
-i\eps_{IJKL}e^{J}\wedge e^{K}\wedge e^{L}\wedge\big(D^{(\cal A^{+})}\chi^{*A}\sigma^{I}_{AA'}\chi^{A'}\big) \label{tospinors2}
\end{align}
It is the latter possibility that was considered by Ashtekar et al. \cite{Ashtekar:1989ju} in the self-dual Einstein Cartan theory where the field ${\cal A}^{-IJ}$ does not appear in the formalism and hence the term (\ref{tospinors1}) cannot be constructed. It was shown that the coupling (\ref{tospinors2}) nonetheless allowed the recovery of familiar results from the coupling of gravity to spinors in Einstein-Cartan theory. In the present model, if the coupling (\ref{tospinors2}) is chosen then the spinor field does not `see' the curvature of the background. If, on the other hand, the coupling (\ref{tospinors1}) is chosen then a brief calculation shows that if $\chi^{A'}$ is part of a Dirac spinor $\Psi$ (for example the left handed electron-neutrino in the standard model) then the following couplings appear in the spinor Lagrangian on this background:  $a_{I}\bar{\Psi}\gamma^{I}\Psi$ and $b_{I}\bar{\Psi}\gamma^{5}\gamma^{I}\Psi$ where $a_{I},b_{I} \sim n_{I}/t$. Such couplings have been widely studied in the context of Lorentz-violating extensions of the standard model \cite{Colladay:1998fq} and contemporary constraints \cite{Kostelecky:2008ts,Heckel:2008hw,Ferrari:2018tps}  on the magnitude of components of $b_{I}$ would correspond to a $t$ value of the order of several months.
The fact that $\{a_{I},b_{I}\}$ diverge $t\rightarrow 0$ is perhaps indicative that treating the matter coupling to gravity via a term (\ref{tospinors1}) as a small perturbation to the background Section \ref{curvedflat} is not consistent.
Nonetheless, it is conceivable that the field configuration $\{X^{I},{\cal A}^{IJ}\}$ producing the geometry accessible to experiment approximates a part of this solution and so the above Lorentz-violating matter couplings may be relevant, however a definitive answer likely depends on the resolution of the dark matter propagation issue discussed earlier.

\section{Discussion and conclusions}
\label{discussandconclude}

The aim of this paper was to clarify a novel approach the recovery of gravitational theory via a gauging process. Originally, Kibble improved Utiyama's theory by considering special-relativistic actions and gauging their global Poincar\'{e} invariance, considered as a combination of `internal' Lorentz transformations and a subgroup of the spacetime diffeomorphism group, thus introducing fields $\omega^{IJ}_{\mu}$ and $e^{I}_{\mu}$ that admitted an interpretation as gauge fields. An alternative approach, with a rationale analogous to parameterised Newtonian mechanics, could be regarded as the gauging of some or all of the global Poincar\'{e} symmetry (\ref{xiglob}) of the dynamical fields $X^{I}$. The known recovery of what may be considered standard gravitational theory upon gauging the full Poincar\'{e} group (in the form of the Einstein-Cartan theory of gravity) or via its subgroup of translations (in the form of the teleparallel equivalent to General Relativity) was discussed. 

It was then shown that the gauging of the Lorentz group could yield a different theory: an extension of General Relativity with novel phenomenology, notably the existence of a modification to Einstein's equations interpretable as a dark matter component\footnote{An interesting alternative modification to gravity has been considered \cite{Anderson:1999wz} in the \emph{metric} formalism  inspired by the step from Newtonian mechanics to parameterized particle mechanics and also resulting in an effective dark matter component.}. The gauging of translations in the gauge theory based on the Lorentz group is completely different to that in the Poincar\'{e} and teleparallel case. No affine generalisation of the Minkowski space is required, and no $\theta^{I}_{\mu}$ is introduced to localise translations.

It is notable that the gauging of Lorentz symmetry implies that the local Lorentz symmetry must be a \emph{complex} one, suggesting the possibility that gravitational fields may be complex in some circumstances (though we assume that spacetime coordinates are real throughout). Preliminary results suggest that this allows for solutions corresponding to metrics of Euclidean signature to exist. Exploration of general solutions to the theory possessing real spacetime metric will be aided by the canonical formulation of the action (\ref{loract}) \cite{NikjooZlosnik23}. Do there exist classical solutions within this model that allow for the dynamical signature change of the four dimensional metric?
	
It may additionally be asked whether the action (\ref{loract}) is somehow preferred, beyond the parameter choice $\beta= \pm i$. It may be shown \cite{Gallagher:2021tgx} that up a boundary term it is equal to:

\begin{align}
S &= \alpha\int \big(\frac{1}{4}X^{2}\eps_{IJKL}  +2\beta  X_{J} X_{L} \eta_{IK}\big) R^{IJ}\wedge R^{KL} \label{rxrx}
\end{align}
where here $R^{IJ}=R^{IJ}({\cal A})$ i.e. the curvature of the full spin connection. One additional term that is possible that is both quadratic in $X^{I}$ and in $R^{IJ}$:  $X^{2}R_{IJ}\wedge R^{IJ}$; it is known \cite{Zlosnik:2018qvg} that unless $\beta=\pm i$ in (\ref{rxrx}) then General Relativity is not recovered, instead yielding a modified theory where, for example, gravitational waves do not travel at the speed of light. It is conceivable that this putative additional term in the Lagrangian causes similar effects. Additionally, if it is required that actions are invariant (up to a boundary term) under covariant-constant translation symmetry (i.e. transformations $X^{I} \rightarrow X^{I} + \phi^{I}$ that satisfy $\dlo \phi^{I}=0$) then this term and terms with arbitrary powers of $X_{I}X^{I}$ are excluded.

It was shown that the action (\ref{loract}) possesses solutions corresponding to General Relativity coupled to a pressureless perfect fluid \cite{Brown:1994py}. It is tempting to wonder whether this fluid could be responsible for some or all of the dark matter. In Section \ref{section_phenomenology} we have discussed issues facing the model as a viable origin for dark matter and different possibilities for their resolution, notably either via quantum corrections to the model or the possibility of a `UV completion' of the model where the effect of new degrees of freedom beyond those resulting from (\ref{loract}) become important.

Though in this paper we restricted to the (inhomogeneous and homogeneous) Lorentz groups, the new approach to translations can be applied in the more general context of gravitational gauge theories. For example, metric-affine gravity is conventionally formulated on the general affine bundle, but for local translation invariance the general linear bundle would be sufficient (see also \cite{Koivisto:2019ejt}). A generic consequence of the latter approach is that torsion, if defined as $D^2 X^I$, is not independent of the curvature but equal to $R^{I}_{\ph{I}J} X^J$. Our conclusion that a flat metric may then correspond to non-trivial gauge geometry in terms of curvature and torsion, seems to be at odds with the
	conventional interpretation that in the absence of gravity $g_{\mu\nu}=\eta_{\mu\nu}$. However, the more natural ground state is rather $g_{\mu\nu}=0$, if the metric is a composite field 
	as it is considered e.g. in the contexts of emergent spinor gravity \cite{Akama:1978pg,Diakonov:2011im,Wetterich:2021ywr,Wetterich:2021hru,Gallagher:2022kvv}.
	Whereas the conventional approach to gauging external symmetries endows a spacetime with gravitational interaction, the novel approach better describes the emergence of 
	a spacetime in concert with gravity.

\paragraph{Acknowledgements: } We thank Amel Durakovic, Priidik Gallagher, Sean Gryb, Pavel Jirou\v{s}ek, Luca Marzola, Sabir Ramazanov, Alexander Vikman, and Hans Westman for helpful discussions. TZ acknowledges support from the Grant Agency of the Czech Republic
GACR grant 20-28525S. This research is part of the project No. 2021/43/P/ST2/02141 co-funded by the Polish National Science Centre and the European Union Framework Programme for Research and Innovation Horizon 2020 under the Marie Sk\l{}odowska-Curie grant agreement No. 945339. TK acknowledges support from the Estonian Research Council grants PRG356 and MOBTT86, and from the European Regional Development Fund CoE program TK133.

\appendix

\section{Matrix representation of the Poincar\'{e} group}
\label{poin_matrix}
The Poincar\'{e} group admits a five dimensional matrix representation, with a group element having the form:

\begin{align}
P^{A}_{\ph{A}B} &= \begin{pmatrix} 
\Lambda^{I}_{\ph{I}J} & \xi^{I}  \\
0 & 1 \\
\end{pmatrix}
\end{align}
and the following generators:

\begin{align}
({\cal P}^{I}_{\ph{I}J})^{A}_{\ph{A}B}&= \begin{pmatrix} 
j^{I}_{\ph{I}J} & 0 \\
0 & 0 \\
\end{pmatrix} \\
({\cal P}^{I})^{A}_{\ph{A}B} &= \begin{pmatrix} 
0 & p^{I} \\
0 & 0 \\
\end{pmatrix} \\
\end{align}
where the sub-matrices $(j^{IJ})^{K}_{\ph{K}L}$ are the generators of the Lorentz group and e.g. $p^{0}=(1,0,0,0)$. It can be shown that the matrices $\{{\cal J}^{IJ},{\cal P}^{I}\}$ indeed satisfy the Lie algebra of the Poincar\'{e} group and so a connection ${\cal P}^{A}_{\ph{A}B\mu}$ in this Lie algebra can be written:

\begin{align}
{\cal P}^{A}_{\ph{A}B\mu} &= \begin{pmatrix} 
\omega^{I}_{\ph{I}J\mu}& \theta^{I}_{\mu}  \\
0 & 0 \\
\end{pmatrix}
\end{align}
Under a gauge transformation represented by a matrix $P^{A}_{\ph{A}B}$ we require that:

\begin{align}
{\cal P}^{A}_{\ph{A}B\mu} &\rightarrow P^{A}_{\ph{A}C}{\cal P}^{C}_{\ph{C}D\mu}(P^{-1})^{D}_{\ph{D}B} - \partial_{\mu}P^{A}_{\ph{A}C}(P^{-1})^{C}_{\ph{C}B}
\end{align}
An object in the fundamental representation of the group is here taken to be a five-vector $X^{A} = (X^{I},1)$ with covariant derivative

\begin{align}
D^{({\cal P})}_{\mu}X^{A} = \partial_{\mu}X^{A}+ {\cal P}^{A}_{\ph{A}B\mu}X^{B}
\end{align}
It can be checked that the $I$ components of this equation are indeed equal to (\ref{poin_codiv}) whilst the `4' component is simply equal to zero.

\section{Illustration of vector field profiles for zero curvature and non-zero curvature Minkowski solutions}
\label{thefigure}

\begin{widetext}
	\begin{figure}[h!]
		\begin{center}
			\includegraphics[]{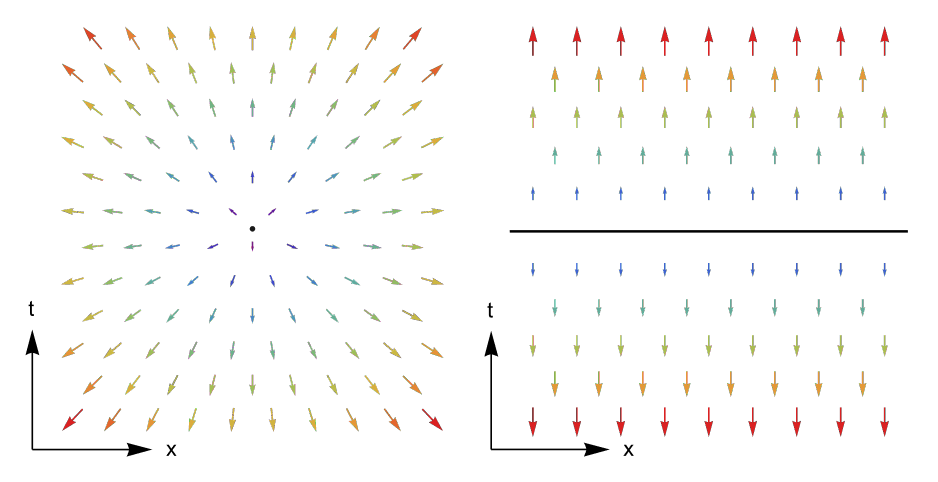}
			\caption{Profiles for $X^{I}$ in a plane coordinatized by inertial coordinates $(x,t)$ leading to Minkowski metric with zero (left panel) and non-zero (right panel) spacetime curvature respectively. Points where $X^{I}=0$ are given in black.}
		\end{center}
	\end{figure}
\end{widetext}

\clearpage
\bibliography{references}

\end{document}